\documentclass[3p,onecolumn,11pt]{revtex4-2}

\usepackage{layouts}
\usepackage{graphicx}
\usepackage{amssymb}
\usepackage{float} %added by Javier for the first iteration

\usepackage{soul}
\usepackage{lineno}

\usepackage{hyperref}
\usepackage[dvipsnames]{xcolor}

\graphicspath{{figures/}}

\usepackage{setspace}

\usepackage{appendix}
\usepackage{bm}
\usepackage{amsmath}
\usepackage{amsfonts}
\usepackage{mathrsfs}
\usepackage{comment}
\usepackage{color}
\usepackage{hyperref}
\hypersetup{
    colorlinks=true,
    allcolors=NavyBlue
}
\usepackage{tikz}

\usepackage[dvipsnames]{xcolor}

\renewcommand{\eqref}[1]{Eq.~(\ref{#1})}

\usepackage{lineno}

\begin{document}
\setlength{\footskip}{2cm}


\title{Frictional sliding strength of knotted and capstan configurations \\ along the axis of a cylinder \\ }



% Define Authors and Affiliations
\author{Javier Sabater}
\affiliation{Flexible Structures Laboratory, Institute of Mechanical Engineering, \\ \'Ecole polytechnique f\'ed\'erale de Lausanne, 1015 Lausanne, Switzerland}

\author{Ji-Sung Park}
\affiliation{Flexible Structures Laboratory, Institute of Mechanical Engineering, \\ \'Ecole polytechnique f\'ed\'erale de Lausanne, 1015 Lausanne, Switzerland}

\author{Jérôme Crassous}
\affiliation{University of Rennes, CNRS, IPR (Institut de Physique de Rennes) - UMR 6251, F-35000 Rennes, France}
\affiliation{PMMH, CNRS, ESPCI Paris, Université PSL, Sorbonne Université, Université Paris Cité, F-75005 Paris, France} 

\author{Sébastien Neukirch}
\affiliation{Sorbonne Université, CNRS, Institut Jean le Rond d'Alembert, F-75005 Paris, France}

\author{Pedro M. Reis}
\email{pedro.reis@epfl.ch} % This automatically flags him as corresponding author
\affiliation{Flexible Structures Laboratory, Institute of Mechanical Engineering, \\ \'Ecole polytechnique f\'ed\'erale de Lausanne, 1015 Lausanne, Switzerland}

\begin{abstract}
We investigate the sliding strength of thin filaments in frictional contact with a translating cylinder, perpendicular to the filaments' axes, in knotted (clove hitch) and unknotted (capstan) configurations. Recent work reported superlinear scaling for surgical knots with elasto-plastic filaments \cite{johanns_strength_2023}. Testing the clove hitch with various materials (elastomeric rods, metallic wires, braided ropes) reveals similar nonlinear behavior, ruling out plasticity. To explore the source of the previously reported nonlinear behavior, we perform three-dimensional FEM simulations (resolving full 3D mechanics) and reduced-order DER simulations (isolating geometric effects by neglecting cross-sectional deformation). Both FEM and DER simulations reproduce the experimental scaling. Simplifying the knot topology by studying capstan angles from $\pi/4$ to $4\pi$ yields comparable superlinear behavior, transitioning to linearity at smaller angles. We rationalize the results by developing an analytical model based on planar \textit{elastica} theory for the capstan configuration (which exhibits behavior similar to the clove hitch but with a simpler topology). The model reproduces the observed superlinear behavior and rationalizes it by coupling the evolution of normal forces and contact arclength during tightening. The analysis further predicts transition to linearity when full contact between the filament and the cylinder is established, providing a mechanical framework applicable across materials, geometries, and topologies.
\end{abstract}

\maketitle

\clearpage

\section{Introduction}
\label{sec:introduction}

Flexible filaments, rods, wires, and ropes in frictional contact with other objects form the mechanical basis of structures across scales and contexts, from climbing plants \cite{silk2005importance, goriely2006mechanics} and bird nests \cite{weiner2020mechanics}, to rigging systems in sailing and climbing \cite{day1986art, wright1928knots}, engineering belts \cite{baser2010theoretical, childs2019belt}, textiles \cite{seguin2022twist, warren2018clothes, poincloux2018crackling, bueno2019structure}, and medical sutures \cite{zimmer1991influence, kim2001significance}. Important aspects of the mechanics of these systems emerge from the intricate coupling between topology, geometry, elasticity, and friction~\cite{grandgeorge_mechanics_2021}, governing how forces are transmitted, resisted, or redistributed through contact. Despite their ubiquity, predicting the mechanical behavior of such systems remains challenging due to
nonlinear geometry, evolving contact regions, and the interplay between bending, tension, and friction~\cite{grandgeorge_elastic_2022}. Studying simplified configurations as model systems, such as filaments wrapped around rigid bodies or self-contacting knots~\cite{grandgeorge_mechanics_2021,grandgeorge_elastic_2022,Singh2022}, provides essential building blocks for addressing more complex problems in this domain.

A canonical problem in the class of systems mentioned above is the capstan configuration, where a filament wraps around a rigid disk. The classical capstan equation, derived by Euler in 1769~\cite{euler_remarques_1769,eytelwein1842handbuch}, predicts that the tension profile along an inextensible filament with negligible thickness and zero bending stiffness is exponential along the contact arc: $T_1 = T_2 e^{\mu \phi}$, where $T_1$ and $T_2$ are the tensions at the two ends, $\mu$ is the friction coefficient between filament and disk, and $\phi$ is the wrap angle. Recent work has extended this framework to elastic filaments with finite cross-sections~\cite{stuart_capstan_1961,jung_capstan_2008,grandgeorge_elastic_2022,Singh2022}, revealing that bending stiffness and finite thickness alter the extent of contact and force transmission along the filament. However, these investigations focus on \textit{tangential} force transmission, addressing the classic question of how forces differ at the two rod ends. Our focus is different: we investigate the force required to translate the cylinder \textit{axially}, perpendicular to the wrapping plane. We define \textit{sliding} to be this perpendicular translation of the cylinder through the wrapped filament.

Beyond capstan configurations, physical knots introduce additional complexity through topology, 3D geometry, and frictional self-contact. Analytical models based on Kirchhoff rod theory have progressed from loose to progressively tighter configurations~\cite{audoly_elastic_2007,jawed_untangling_2015,johanns_shapes_2021,grandgeorge_mechanics_2021,sano_exploring_2022}, while fully 3D finite-element modeling (FEM) studies have resolved the strong curvature and tight contact that arise in highly bent knots~\cite{baek_finite_2020}. Recent work has further shown how friction governs stability, including twist-mediated capsizing in stopper knots~\cite{johanns_capsizing_2024} and tension-drop driven locking units in the bowline~\cite{aymon2025self}. Despite these advances, there are still many open questions. Among the numerous knot types~\cite{ashley_ashley_1944}, sliding knots represent a special class characterized by their ability to be displaced to a desired location and subsequently tightened by pulling on the filament ends. These knots consist of half-hitches wrapped around a tensioned central segment, with the clove hitch (formed by two identical throws in the same direction) serving as the fundamental structural unit. Remarkably, the clove hitch is topologically equivalent to the flat knots commonly used to tie shoelaces, specifically the square and granny knots, differing only in their crossing patterns and spatial configuration. In a parallel investigation into the wrapping mechanics of heavy elastic rods, \citet{tani2024} combined experiments with Kirchhoff rod theory to characterize the transition between tight coiling and helical wrapping configurations formed around a rotating cylinder, demonstrating that the resulting configuration is governed by the interplay of bending elasticity, gravity, and frictional contact.

Recently, \citet{johanns_strength_2023} investigated the mechanics of such surgical sliding knots using polymeric (elasto-plastic) filaments, combining experiments with FEM simulations. They reported data consistent with a striking superlinear relationship between the sliding strength $F_0$ (the force required to steadily pull the cylinder through the knot) and the tying pre-tension $T$. The authors claimed consistency with a power-law $F_0 \sim K \, T^\alpha$ where $\alpha = 1.59 \pm 0.03$ and where the prefactor $K$ depends on the number of throws and friction coefficient.
This scaling was robust across different topologies and validated through FEM simulations. However, the accessible tension range was limited to roughly one order of magnitude, bounded at low tensions by loose knots and at high tensions by filament fracture. Moreover, the polymeric filaments deformed plastically during tightening, with plastic deformation retaining the knot shape upon releasing the pre-tension. Forces were scaled by $\sigma_y \, A$, where $\sigma_y$ is the yield strength and $A$ is the cross-sectional area. These observations raise fundamental questions: \textit{Is plasticity essential for the reported superlinear behavior? Does the scaling truly follow a power law across a broader range of tensions? What physical mechanism drives this behavior?}. The answers remain unclear: existing analytical frameworks for elastic filaments in contact do not predict or explain such a nonlinear force scaling with tying tension, and it is unknown whether this behavior requires plastic deformation, depends on knot topology, or persists across broader ranges of constitutive behaviors. A systematic investigation is needed to isolate the essential ingredients and develop a predictive mechanical framework.

Here, we conduct a systematic investigation of the sliding strength of both knotted (clove hitch) and unknotted (capstan) configurations using an elimination strategy. First, we test whether plasticity is essential by studying purely elastic materials (elastomeric rods and superelastic metallic wires) and an inhomogeneous, inelastic braided rope that exhibits irreversible deformation due to internal friction. Unlike the findings of Ref.~\cite{johanns_strength_2023}, where the plastic filaments retained their knot shape even after pre-tension was released, our elastic systems require that the tension ($T$) at both ends be held continuously throughout testing. Second, we set free from the knotted topology by studying unknotted capstan configurations with systematically varied imposed capstan angles, from $\pi/4$ to $4\pi$. Third, we complement the experimental results with both 3D FEM simulations (which resolve full mechanical behavior), and reduced-order discrete element rod (DER) simulations (which isolate geometric effects by neglecting cross-sectional deformation). Across all material systems, configurations, and both computational approaches, we observe consistent superlinear scaling within a specific range of tension, demonstrating that the behavior is remarkably general. Finally, we develop an analytical model, based on planar \textit{elastica} theory for the capstan configuration, which captures the observed nonlinearity and reveals its origin: the coupled evolution of normal contact forces and contact arclength during tightening. The model further predicts a transition to linear scaling at full contact, providing a mechanical framework applicable across materials, geometries, and topologies.

Our paper is organized as follows. Section~\ref{sec:problem_definition} defines the problem, configurations, and nondimensionalization. Sections~\ref{sec:experimental_methodology} and~\ref{sec:methodology_numerics} describe the experimental and numerical methods. Section~\ref{sec:results_CH} presents results for clove hitch and capstan configurations for different material systems, comparing experiments with FEM and DER simulations. Sections~\ref{sec:elastica_model} and~\ref{sec:elastica_validation_results} develop and validate the planar elastica model. Section~\ref{sec:contact} characterizes the evolution of contact geometry. Finally, Section~\ref{sec:conclusions} summarizes and discusses our main findings.

%%%%%%%%%%%%%%%%%%%%%%%%%%%%%%%%%%%%
\section{Definition of the problem}
\label{sec:problem_definition}
%%%%%%%%%%%%%%%%%%%%%%%%%%%%%%%%%%%%

\begin{figure}[t!]
    \centering
    \includegraphics[width=0.7\textwidth]{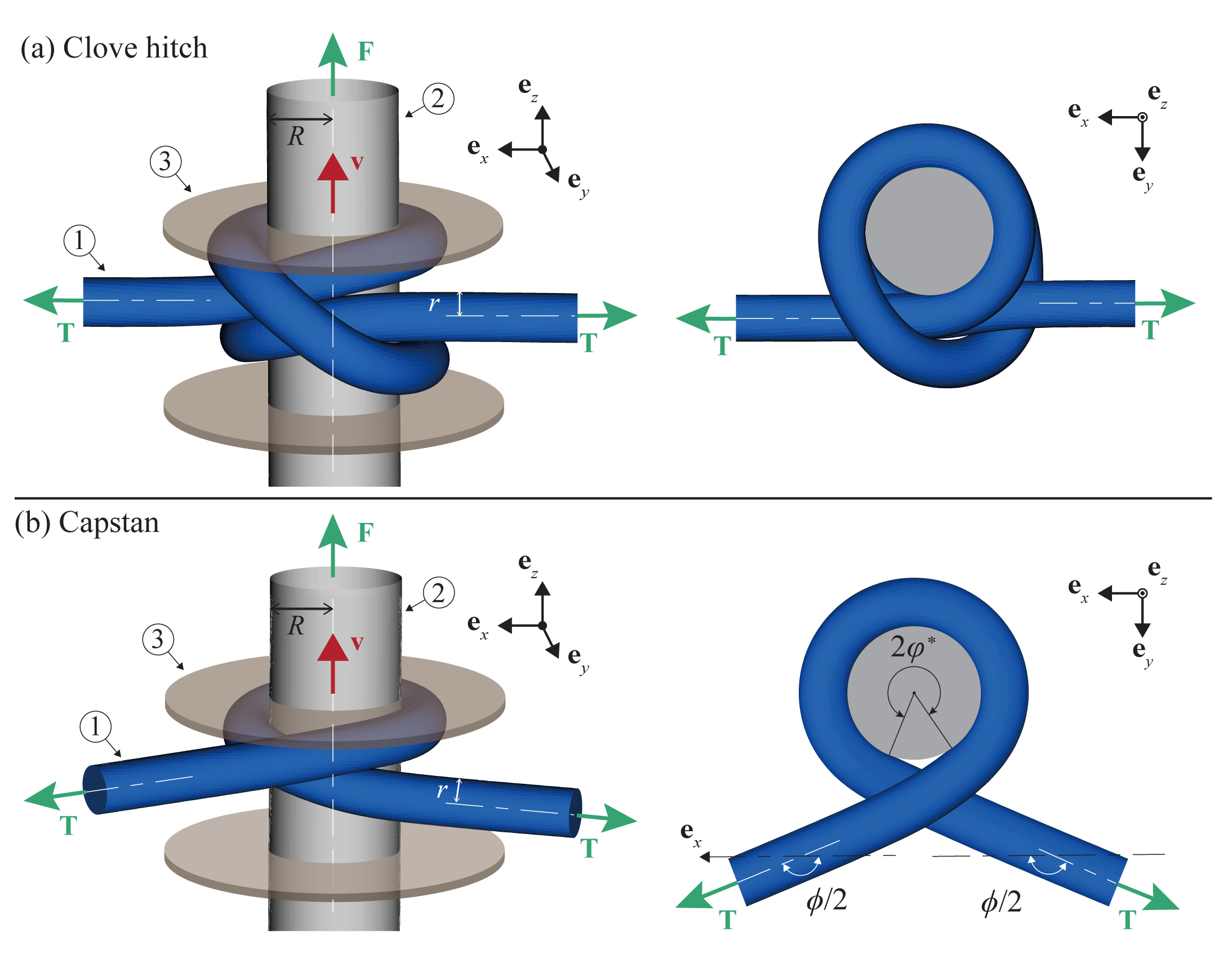}
    \caption{
    Schematics of a flexible filament wrapped around a vertical cylinder for two configurations: (a) clove hitch and (b) capstan; 3D views on the left and top views on the right. For both systems, the filament (1) has circular cross-section of radius $r$ and is subjected to equal tying tensions of magnitude $\mathbf{T}$ at both ends. The cylinder (2) is pulled at a velocity $\mathbf{v}$ (along $\mathbf{e}_z$). The filament is pressed against the stopper plate (3). For the capstan, panel (b) illustrates the angular extent of contact $2\varphi^{*}$ (where $\varphi^{*}$ is the wrapping angle) and the angle $\phi/2$ that the applied tension $\mathbf{T}$ makes with $\mathbf{e}_y$ (where $\phi$ is the capstan angle). We vary $T=\lvert\mathbf{T}\lvert$ and $\phi$ systematically.}
\label{fig:schema_CH_capstan}
\end{figure}

We proceed by describing the problem setup: the clove hitch and capstan configurations, their geometric parameters, and the nondimensionalization that enables quantitative comparison across systems. Figure~\ref{fig:schema_CH_capstan} illustrates the two knotted and unknotted configurations we study: (a) clove hitch and (b) capstan. In both systems, a flexible filament (1) with a circular cross-section of radius $r$ wraps horizontally around a vertical cylinder (2) of radius $R$ and is constrained vertically by two stopper plates (3). Equal tensions $\mathbf{T}$ are applied symmetrically at both filament ends with an angle $\phi/2$ with respect to $\mathbf{e}_x$. We measure the pulling force $F$ required to translate the cylinder at constant velocity $\mathbf{v}$, along $\mathbf{e}_z$, and define the sliding strength $F_0$ as the mean plateau value. Frictional contact occurs between the filament and cylinder surface, and at filament-filament interfaces in the clove hitch. 

For the capstan configuration, the capstan angle $\phi$ (which defines the direction of applied tension) serves as a geometric control parameter that we vary from $\pi/4$ to $4\pi$. Note that in our notation, $\phi$ differs from the wrap angle used in the classical capstan equation; here $\phi$ is an imposed geometric parameter while the contact arc $\varphi^{*}$ is an obtained geometric quantity. As the tension $T$ increases, the contact arc grows, with the wrapping angle $\varphi^{*}$ evolving from near zero (loose wrapping) toward $\phi/2$ (fully tightened configuration). Unlike the classical capstan problem~\cite{euler_remarques_1769, eytelwein1842handbuch}, where a flexible filament wraps around a rigid circular drum with fixed geometry and asymmetric loading, our system involves symmetric loading (equal tensions at both ends), finite bending stiffness, and contact geometry that evolves with tension. Previous studies of elastic filament-cylinder interactions~\cite{grandgeorge_elastic_2022, Singh2022} have addressed tangential force transmissions in the $\mathbf{e}_{x}$-$\mathbf{e}_{y}$ plane. Here, we instead focus on the normal pulling force $F$ required to translate the cylinder in the $\mathbf{e}_{z}$ direction through the wrapping.

Addressing our research questions quantitatively across systems with different materials and geometries requires appropriate nondimensionalization. \citet{johanns_strength_2023} nondimensionalized forces by $\sigma_y \, A$, where $\sigma_y$ is the yield strength and $A$ the cross-sectional area of the filament. This was a natural choice for plastically deforming filaments. However, for elastic filaments (and ropes with an effective bending stiffness), the relevant force scale arises from bending deformation of the wrapped filament: $EI/(R+r)^2$, where $EI$ is the bending stiffness, and we recall that $R$ and $r$ are the radii of the cylinder and filament, respectively. We use this scale to nondimensionalize all forces: $\tilde{T}=T(R+r)^2/(EI)$, $\tilde{F}=F(R+r)^2/(EI)$, and $\tilde{F}_0=F_0(R+r)^2/(EI)$.

%%%%%%%%%%%%%%%%%%%%%%%%%%%%%%%%%%%%
\section{Methodology: Experiments}
\label{sec:experimental_methodology}
%%%%%%%%%%%%%%%%%%%%%%%%%%%%%%%%%%%%

\begin{figure}[h!]
    \centering
    \includegraphics[width=0.9\textwidth]{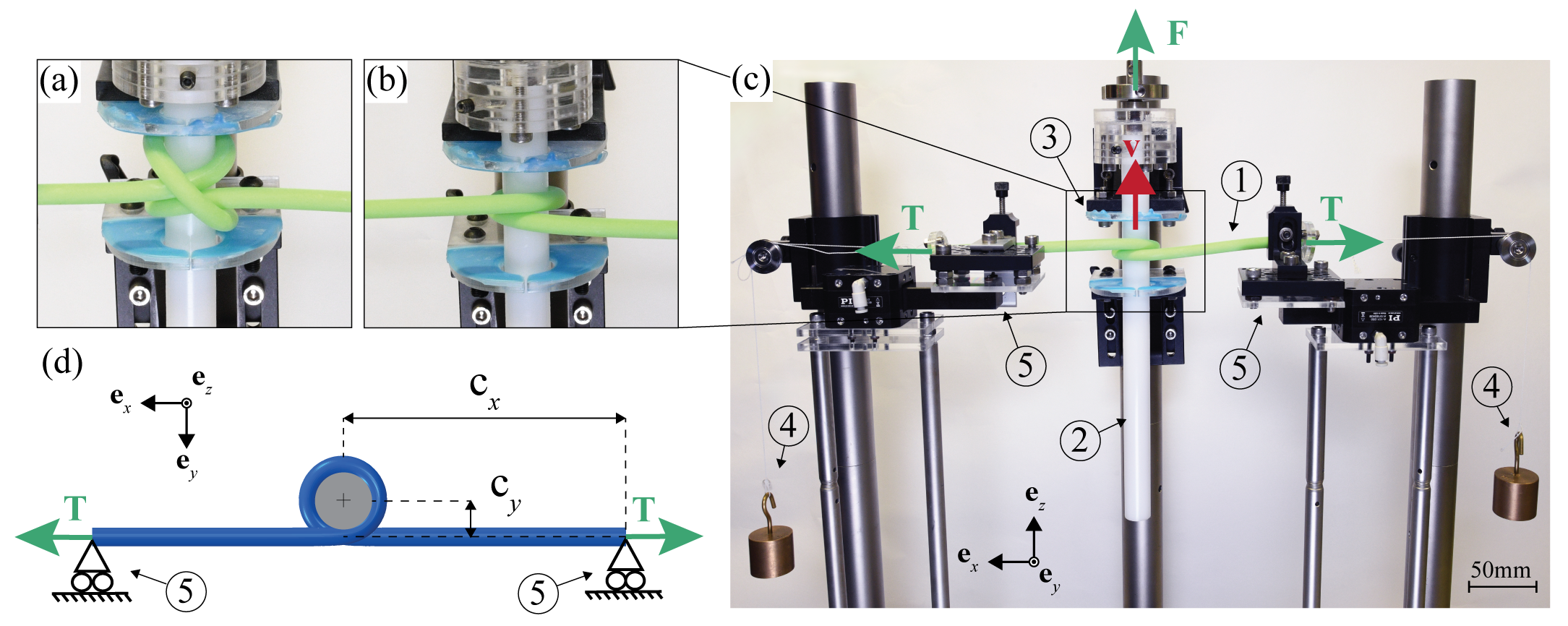}
    \caption{Experimental system. (a) Clove hitch and (b) capstan ($\phi=2\pi$) configurations with VPS rod around POM cylinder. (c) Full experimental apparatus. (d) Schematic of boundary conditions at $(\pm c_x,\,c_y)$ from the cylinder center.}
    \label{fig:experimental_apparatus}
\end{figure}

We conducted experiments with the clove hitch and capstan configurations (Fig.~\ref{fig:schema_CH_capstan}) using elastomeric rods, metallic wires, and braided ropes wrapped around an identical filament or a rigid cylinder. The photographs in Fig.~\ref{fig:experimental_apparatus}(a, b) show a representative combination (a VPS rod around a POM cylinder) for both the clove hitch and capstan configurations in the experimental apparatus. These experiments investigate the generality of the nonlinear scaling observed by \citet{johanns_strength_2023} across different material systems and topologies. In what follows, we describe the apparatus, material systems, and protocols for measuring the sliding strength $F_0$.

\noindent \textit{Apparatus:} Figure~\ref{fig:experimental_apparatus}(c) shows the experimental apparatus for measuring the sliding strength $F_0$. A flexible filament (1) wraps around a vertical cylinder (2), which we mounted to a Universal Testing Machine (UTM, Instron 5943) for force measurement. Two acrylic stopper plates (3) ($8\,\mathrm{mm}$ thick) constraining the vertical ($\mathbf{e}_z$) displacement of the filament. We applied the tying tension $T$ by hanging equal masses (4) at both ends. To prevent twisting and capsizing, we clamped each end of the filament to a linear slider (5) (air-bearing stage, Physik Instrumente A-10x) at $(\pm c_x,\,c_y) = (\pm 100,\, r+R)\,\mathrm{mm}$ from the cylinder center (Fig.~\ref{fig:experimental_apparatus}d).

\noindent \textit{Material systems:} We tested three filament types in multiple configurations (Table~\ref{tab:Rod_configurations}): elastomeric rods (extensible and inextensible variants), Nitinol wires, and a braided rope, wrapped around either a flexible cylinder of the same material or a rigid cylinder. We cast the elastomeric rods in-house from vinylpolysiloxane (VPS 32, Zhermack Elite Double 32) into cylindrical specimens with a radius of 4.0\,mm and a length of 300\,mm. VPS behaves as an incompressible Neo-Hookean solid ($E = 1.2\,\mathrm{MPa}$, $\nu \approx 0.5$) up to a stretch ratio of $\lambda=1.30$ \cite{baek_finite_2020, grandgeorge_mechanics_2021, johanns_shapes_2021}. We prepared two variants of the VPS rods: extensible (Ext.) and nearly inextensible (Inext.). We reinforced the latter with a central Nitinol wire ($d = 0.127\,\mathrm{mm}$, $E = 67.5\,\mathrm{GPa}$, Dynalloy, Inc.) to effectively suppress axial stretching and minimize cross-sectional deformation \cite{johanns_capsizing_2024}. For both variants, we tested wrapping around a VPS cylinder ($R=r$) or a larger rigid polyoxymethylene (POM) cylinder ($E_{POM} \approx 3\,\mathrm{GPa}$, $R/r=2.5$). We treated all VPS surfaces, including the VPS-coated stopper plates, with talcum powder for reproducible friction~\cite{grandgeorge_mechanics_2021}. For another set of experiments, we wrapped Nitinol wires ($r = \{0.055, 0.10, 0.15\}\,\mathrm{mm}$) around a polished steel cylinder ($R = 6\,\mathrm{mm}$), using air bearings instead of the stopper plates. We also tested a braided climbing rope (inelastic behavior) with radius $r=4.0\,\mathrm{mm}$ and effective bending stiffness $EI = 0.4\times10^{-3}\,\mathrm{Nm^2}$. These various configurations enable systematic investigation of how filament slenderness, stiffness, and contact properties govern sliding strength.

\begin{table}[h!]
\centering
\caption{\label{tab:Rod_configurations} Experimental material systems and configurations.}
\begin{tabular}{ll|cc|c}
\hline
\multicolumn{2}{c|}{Configuration} & \multicolumn{2}{c|}{Radius [mm]} & Friction \\
\cline{1-2} \cline{3-4} \cline{5-5}
Flexible rod ($r$) & Cylinder ($R$) & $r$ & $R$ & $\mu$ \\
\hline
Ext. VPS & Ext. VPS & 4.0 & 4.0 & 0.32 \\ 
Inext. VPS & Inext. VPS & 4.0 & 4.0 & 0.32 \\ 
Ext. VPS & POM & 4.0 & 10.0 & 0.32 \\ 
Rope & Rope & 4.0 & 4.0 & 0.28 \\ 
Nitinol & Steel & \{0.055, 0.10, 0.15\} & 6.0 & 0.14 \\
\hline
\end{tabular}
\end{table}

\begin{figure}[h!]
    \centering
    \includegraphics[width=0.9\textwidth]{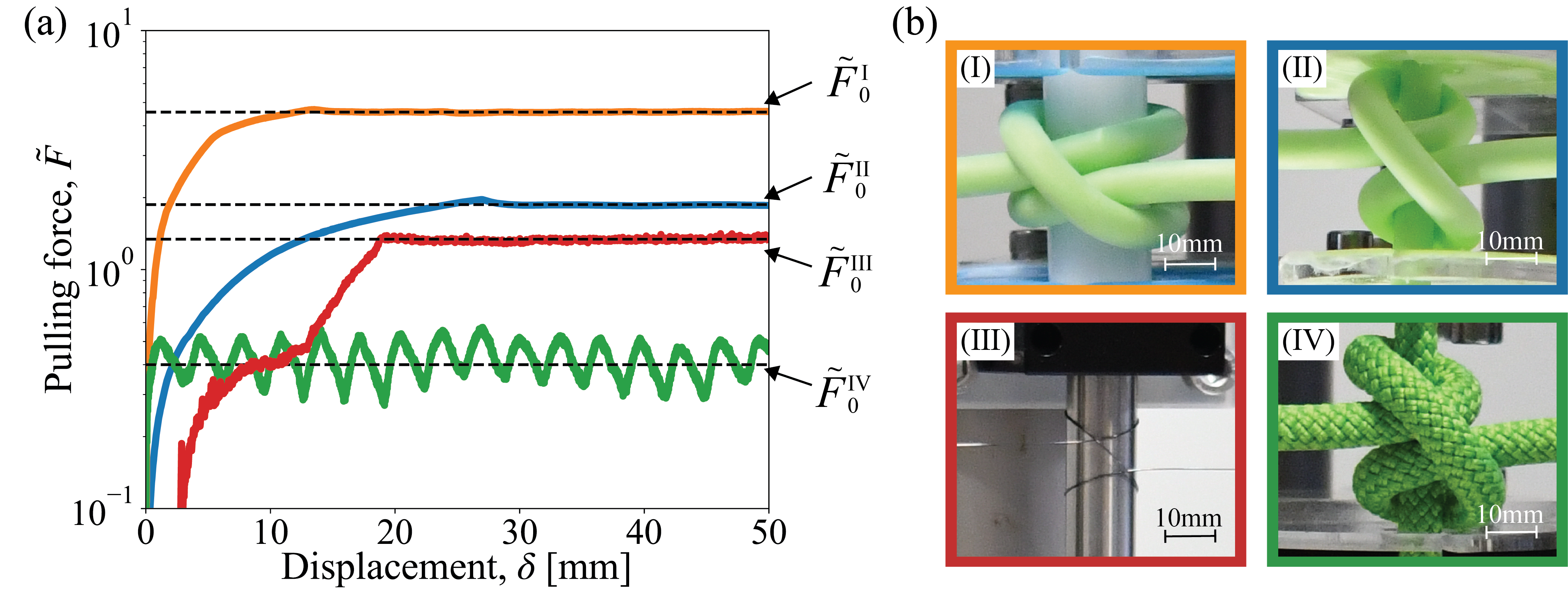}
    \caption{Experimental force-displacement curves for various material systems. (a) Dimensionless pulling force $\tilde{F}$ versus vertical displacement $\delta$ for the configurations shown in (b): ($\mathrm{I}$) Ext. VPS rod around POM cylinder, ($\mathrm{II}$) Ext. VPS rod around VPS cylinder, ($\mathrm{III}$) Nitinol wire around steel cylinder, and ($\mathrm{IV}$) rope around rope. Each curve in (a) shows a representative test at tying tensions $\tilde{T} = \{0.15, 0.5, 0.09, 0.09\}$ for panel (b) $\mathrm{I}$, $\mathrm{II}$, $\mathrm{III}$, and $\mathrm{IV}$, respectively. The dashed line indicates the sliding strength $\tilde{F}_0$ (mean plateau value) at an imposed velocity of $v = \lvert\textbf{v}\lvert=1\,\mathrm{mm/s}$.}
    \label{fig:experimental_protocol}
\end{figure}

\noindent \textit{Experimental protocol:} We assembled each configuration (clove hitch or capstan) according to Table~\ref{tab:Rod_configurations} and applied the tying tension $T$ by hanging identical masses ($50\,\mathrm{g}$ to $10\,\mathrm{kg}$) at both filament ends. We pulled the vertical cylinder upward along $\mathbf{e}_z$ at $v = \lvert\textbf{v}\lvert=1\,\mathrm{mm/s}$ using the UTM, which recorded the pulling force as a function of the imposed displacement. Figure~\ref{fig:experimental_protocol}(a) plots the measured dimensionless pulling force $\tilde{F}$ as a function of vertical displacement $\delta$ for the configurations $\mathrm{I}$, $\mathrm{II}$, $\mathrm{III}$, and $\mathrm{IV}$ shown in Fig.~\ref{fig:experimental_protocol}(b). Each curve represents a single experimental run. Initially, $\tilde{F}$ increases as the filament presses against the upper stopper plate. The cylinder continues to ascend while the wrapped filament remains blocked, creating a brief transient before the force stabilizes to a plateau corresponding to steady sliding. We define the dimensionless sliding strength, $\tilde{F}_0$, as the mean plateau value.

%%%%%%%%%%%%%%%%%%%%%%%%%%%%%%%%%%%%
\section{Methodology: Numerical simulations}
\label{sec:methodology_numerics}
%%%%%%%%%%%%%%%%%%%%%%%%%%%%%%%%%%%%

We performed numerical simulations using both the Finite Element Method (FEM) and the Discrete Element Rod (DER) for the two configurations of interest: clove hitch and the capstan configurations (Fig.~\ref{fig:schema_CH_capstan}). Both methods simulate a VPS rod wrapped around a rigid POM cylinder. For the capstan, we systematically varied the capstan angle. These simulations complement the experiments by providing access to quantities that cannot be directly measured, such as internal stress distributions within the rods and the surface tractions at the contact interfaces. These quantities offer key physical insight into the mechanics governing sliding strength. Next, we describe the FEM model and numerical knotting procedure, followed by the DER model.

\paragraph{FEM simulations} We performed simulations using ABAQUS/Standard 2023 under quasi-static conditions to simulate steady sliding. Contact interactions were modeled using hard contact with penalty enforcement to handle nonlinear interactions as the wrapping rod tightens. The elastic rod ( for the capstan, length $(r+R) \phi + 100$\,mm and radius, $r=4.0$mm  was discretized using reduced-integration hybrid 3D hexahedral elements (\textsc{C3D8RH}) with approximately 200 elements on the cross-section to capture bending. The rigid cylinder (radius $R=10$mm) and stopper plates (outer radius 24.4\,mm, separation 92\,mm) were modeled as discrete rigid shells, with the cylinder meshed at half the rod element size for smooth contact resolution. For the clove hitch, we generated the geometry using a custom displacement subroutine (\texttt{DISP}) based on the mapping technique developed by~\citet{aymon2025self}. The subroutine prescribes the rod-centerline displacement from a straight configuration to the target knot shape obtained from microtomography scans of a physical clove hitch, while minimizing axial stretch. Contact interactions are deactivated during this stage, as the rod temporarily intersects the cylinder. For the capstan, we obtained the wrapping topology by inducing buckling in a straight rod through axial compression along $\mathbf{e}_x$. The rod ends were kinematically coupled to control nodes. After generating the geometry, we released all boundary conditions except at the rod ends to dissipate residual kinetic energy, then applied the tying tension \textbf{T} at both ends. We pulled the cylinder upward at speed $v=1\,\mathrm{mm/s}$ along $\mathbf{e}_z$, with fixed rod-end rotations. Just as in the experiments, after a transient, the pulling force $F$ stabilized to a plateau: the sliding strength $F_0$, see also Section~\ref{sec:experimental_methodology}.

\paragraph{DER simulations} We also performed simulations using a discrete elastic rod model described in detail in Ref.~\cite{crassous_discrete_2023}. The rods consist of articulated segments with circular cross-sections. Elastic forces are determined from bending, torsion, and stretching energies written in discrete form following Ref.~\cite{bergou.2008}, yielding discrete forces and moments acting on the segments. Contacts between segments are modeled as point contacts using the Cundall \& Strack approach~\cite{cundall.1979}: normal forces ${\bf f}_n$ are proportional to the interpenetration, and tangential forces ${\bf f}_t$ to relative displacement, with tangential deformation bounded by Coulomb's condition (${\bf f}_t \le \mu {\bf f}_n$). We integrated Newton's equations of motion using a Verlet algorithm. The simulation consists of two rods: a non-deformable vertical cylinder (translating along its axis) and a wrapping rod discretized into $N$ segments of circular cross-section of radius $r$ and length $l_0 = L/N = r/2$ with material properties $E$ and $\nu=1/2$. The wrapping rod is confined between two large spherical boundaries of radius $10^3 r$ to avoid plane-cylinder contacts. We initialized the wrapping rod with two end segments at an angle $\phi/2$ with respect to $\mathbf{e}_x$, connected tangentially to the wrapped segments of desired configurations: a capstan or a clove hitch. We then applied tensions at the ends and relaxed the system. During relaxation, the central cylinder remained immobile; we set the friction coefficients to zero and allowed the rod to rotate around its axis, ensuring no initial torsion. At equilibrium, we activated friction to nominal values, blocked end rotations, and pulled the central rod along $\mathbf{e}_z$ at $v=1\,\mathrm{mm/s}$. The sliding strength $F_0$ was determined as the mean value of the plateau, as described above. The DER implementation assumes undeformable cross-sections, thereby neglecting the elastic (Hertzian) deformation at the local contacts \cite{hertz1881contact}.

%%%%%%%%%%%%%%%%%%%%%%%%%%%%%%%%%%%%
\section{Sliding strength: experiments and numerical validation} 
\label{sec:results_CH}
%%%%%%%%%%%%%%%%%%%%%%%%%%%%%%%%%%%%

\begin{figure}[h!]
    \centering
    \includegraphics[width=1.0\textwidth]{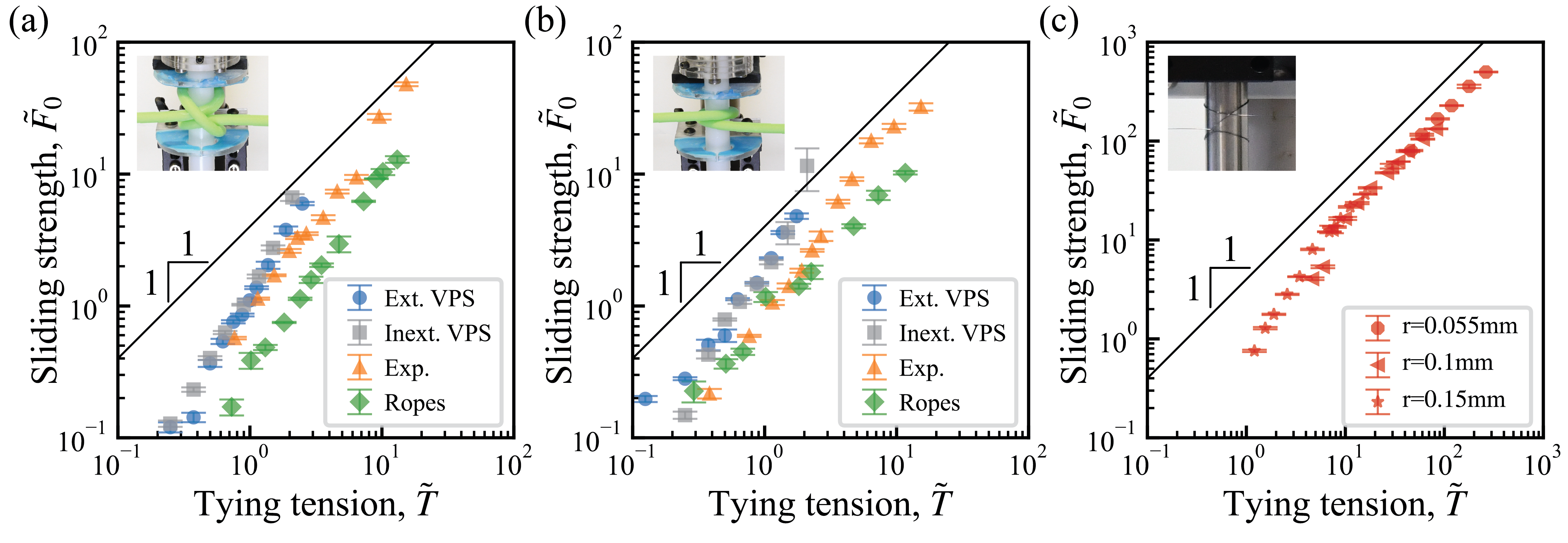}
    \caption{Experimental results of sliding strength for clove hitch and single-loop capstan. (a) Clove hitch: dimensionless sliding strength $\tilde{F}_0$ versus tying tension $\tilde{T}$ for four material systems (see Table~\ref{tab:Rod_configurations}): VPS rod (Ext. and Inext. variants), Ext. VPS rod around POM cylinder, and braided rope. (b) Single-loop capstan ($\phi=2\pi$) for the same material systems. (c) Clove hitch for Nitinol wire of three radii wrapped around a steel cylinder. All configurations exhibit superlinear scaling at low $\tilde{T}$, transitioning to linear behavior at higher $\tilde{T}$. Insets show photographs of the experimental setups. Error bars denote the standard deviation over five measurements. The solid line has unit slope in log-log scale, corresponding to linear scaling ($\tilde{F}_0 \sim \tilde{T}$), evidencing that all data sets exhibit superlinear behaviour in the low $T$ regime.}
    \label{fig:results_CH_2PI}
\end{figure}

\begin{figure}[h!]
    \centering
    \includegraphics[width=1.0\textwidth]{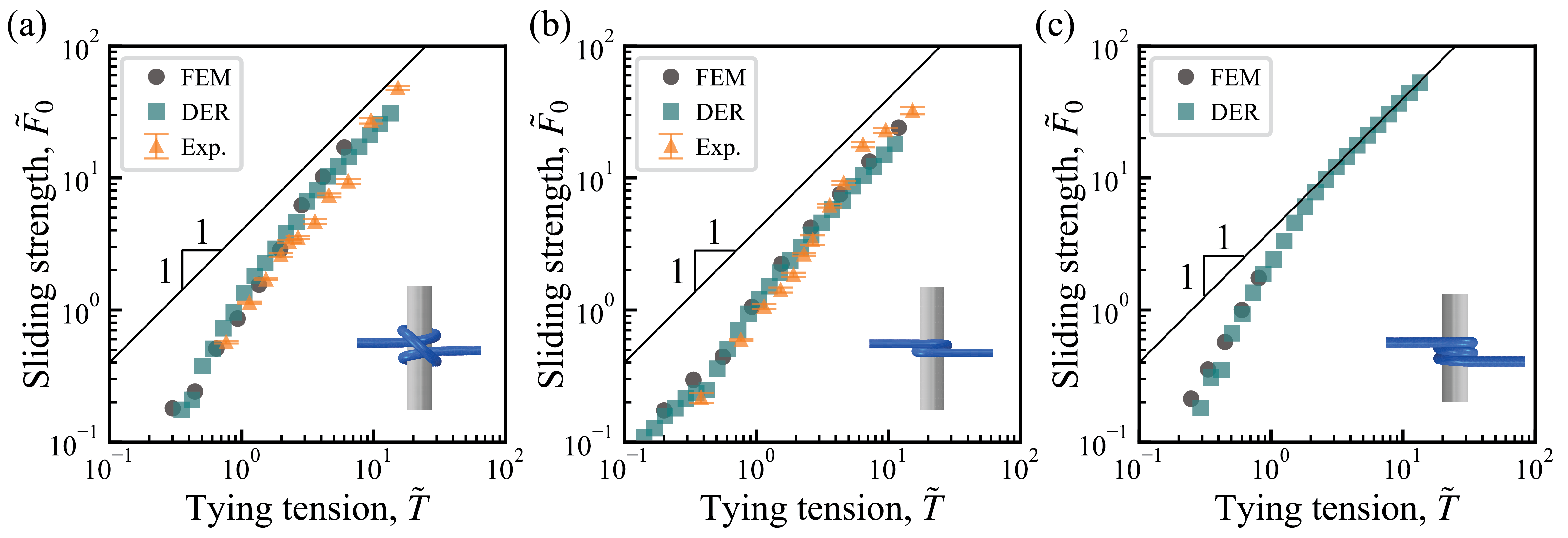}
    \caption{Sliding strength for clove hitch and capstan configurations of an Ext. VPS rod around a POM cylinder: experimental measurements compared with FEM and DER simulations. (a) Clove hitch (b) $\phi=2\pi$ capstan, and (c) $\phi=4\pi$ capstan. Both numerical methods show excellent agreement with experiments. Insets show visualizations obtained from simulations. The solid line has unit slope in log-log scale, corresponding to linear scaling ($\tilde{F}_0 \sim \tilde{T}$).}
    \label{fig:results_FEM}
\end{figure}

Using the experimental and numerical methods introduced in Sections~\ref{sec:experimental_methodology} and~\ref{sec:methodology_numerics}, we present results for the two configurations shown schematically in Fig.~\ref{fig:schema_CH_capstan}, the clove hitch and single-loop capstan ($\phi=2\pi$), across five material systems. All of our data sets plotted in Fig.~\ref{fig:results_CH_2PI} exhibit a superlinear scaling similar to that observed by \citet{johanns_strength_2023}, albeit now also for purely elastic filaments and ropes. We validate our numerical methods by comparing experiments with FEM and DER simulations for a representative system (Fig.~\ref{fig:results_FEM}).

Figure~\ref{fig:results_CH_2PI} shows the dimensionless sliding strength $\tilde{F}_0$ versus tying tension $\tilde{T}$ for the clove hitch and single-loop capstan across five material systems (Table~\ref{tab:Rod_configurations}): Ext. and Inext. VPS rods, VPS rods around POM cylinder, Nitinol wire around steel cylinder, and braided rope. All configurations exhibit superlinear scaling at low $\tilde{T}$, transitioning to an approximately linear behavior at higher tensions. \citet{johanns_strength_2023} first reported this superlinear behavior for elasto-plastic surgical filaments; we confirm it across elastomeric (VPS) and inelastic braided systems (Figure~\ref{fig:results_CH_2PI}a,\,b), demonstrating that the mechanism is not governed by polymer plasticity.

Nitinol results are presented separately in Fig.~\ref{fig:results_CH_2PI}(c), as this material permits access to substantially higher non-dimensional tensions and was tested at three distinct radii. The extended range reveals the linear regime more clearly than in the other systems and beyond the limits reported by \citet{johanns_strength_2023}, whose experiments were constrained by filament fracture. In our measurements, the accessible tension range is bounded below by loss of contact at insufficient tension ($T \ll EI/(R+r)^2$) and above by material failure (plastic yield in Nitinol; fracture in VPS). The close agreement between clove-hitch and capstan trends, and the consistency across elastic, metallic, and inelastic materials, indicates that neither knot topology nor plastic deformation is essential. Therefore, the observed superlinear-to-linear transition reflects underlying geometric and elastic effects governing frictional contact.

To validate our numerical methods, we examine one representative system from Fig.~\ref{fig:results_CH_2PI}: Ext. VPS rods around a POM cylinder. In Fig.~\ref{fig:results_FEM}, we compare experimental measurements with FEM and DER simulations for (a) the clove hitch, (b) the $\phi=2\pi$ capstan, and (c) the $\phi=4\pi$ capstan configuration. Both numerical methods exhibit excellent agreement with experiments and reproduce the superlinear-to-linear transition observed across all material systems. Notably, the DER model, which assumes rigid cross-sections and therefore neglects cross-sectional deformation, captures the scaling equally well. This agreement demonstrates that cross-sectional effects are not essential to the mechanism, further narrowing the mechanism to coupling between bending deformation and frictional contact. The validated numerical methods provide access to internal stress distributions and contact tractions that cannot be measured experimentally, enabling a detailed investigation of the contact mechanics underlying the observed scaling.

The experimental and numerical results presented in Fig.~\ref{fig:results_CH_2PI} and \ref{fig:results_FEM} establish that the observed behavior is general across materials, topologies, and numerical approaches. Systematic elimination rules out material plasticity (elastic and inelastic materials behave quantitatively similarly), knot topology (clove hitch and capstan exhibit identical scaling), and cross-sectional deformation (DER simulations with rigid cross-sections capture the behavior). 

In the next section, we develop an analytical model to rationalize the coupling between tying tension, elastic bending, and contact geometry. Given the qualitative similarity between configurations, we focus on the simpler unknotted capstan geometry.

%%%%%%%%%%%%%%%%%%%%%%%%%%%%%%%%%%%%%
\section{Planar elastica model for capstan configurations}
\label{sec:elastica_model}
%%%%%%%%%%%%%%%%%%%%%%%%%%%%%%%%%%%%%

\begin{figure}[h!]
    \centering
    \includegraphics[width=0.7\textwidth]{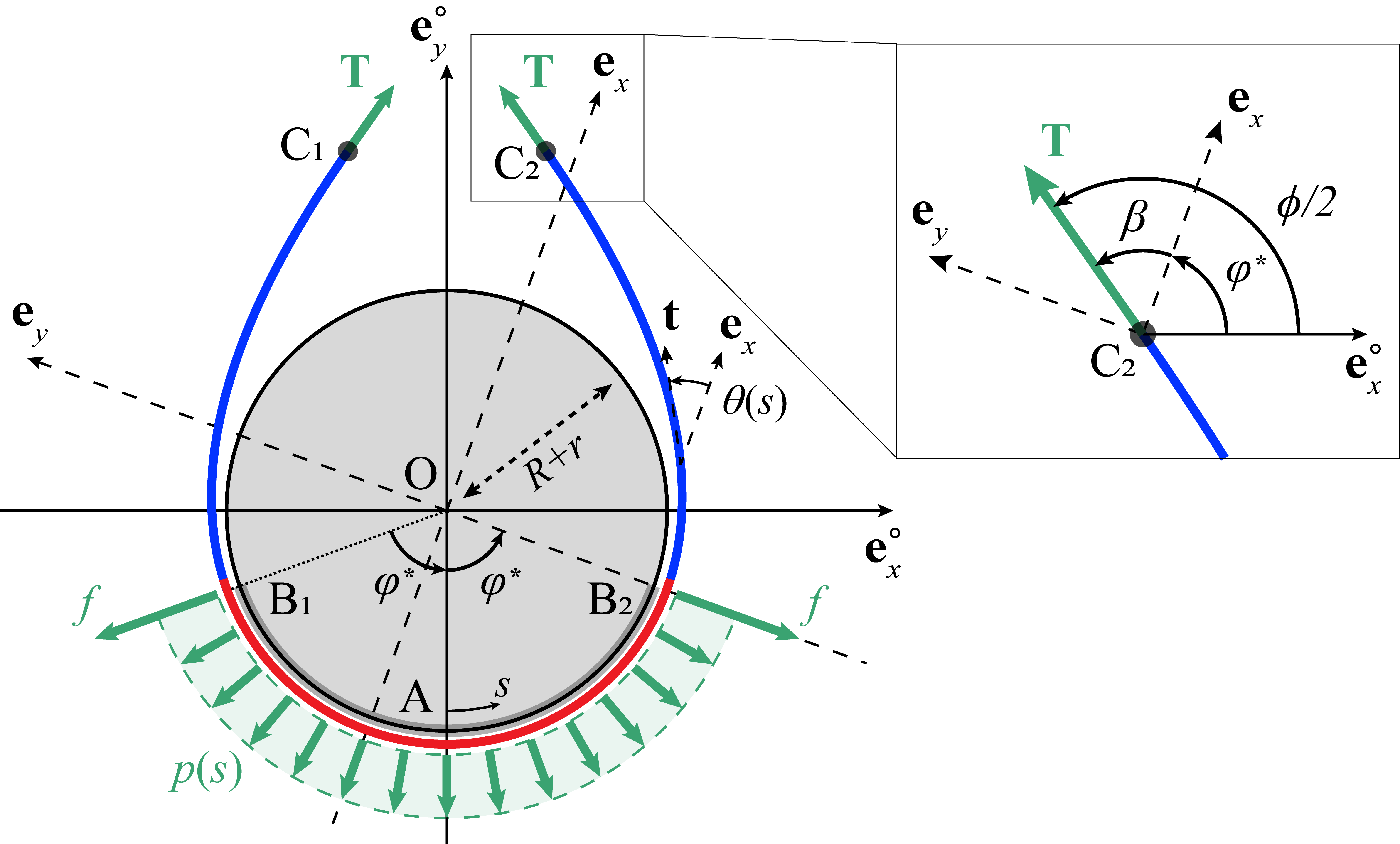}
    \caption{Schematic of the analytical model: a planar elastica wrapped around a rigid disk is pulled by tying tension $\mathbf{T}$ at both ends. Continuous contact occurs between points $B_1$ and $B_2$, with a distributed normal pressure $p(s)$ and point forces $f$ at the contact boundaries. The configuration is symmetric about axis $(OA)$. With no loss of generality, point $B_2$ is chosen to lie at the south pole of $\mathbf{e}_{y}$, so $\theta(s_{B_2})=0$.}
    \label{fig:planar_elastica}
\end{figure}

We rationalize the coupling between tying tension, elastic bending, and contact geometry identified in the previous section, by developing an analytical model based on the planar elastica~\cite{audoly_elasticity_2005,Bigoni2012}. The rod is modeled as an unshearable and inextensible elastic curve of bending stiffness $EI$ wrapped around a rigid disk of radius $R+r$ in a plane perpendicular to the disk's axis (Fig.~\ref{fig:planar_elastica}). This 2D formulation applies to symmetric unknotted configurations, but as we demonstrate later, key predictions extend to the clove hitch. The derivation proceeds in four stages: First, we establish the kinematics and equilibrium of the rod in contact. Second, we utilize Hamiltonian conservation to derive an explicit analytical solution for the total reaction force $R_{\text{tot}}$. Third, we identify the critical tension required for continuous contact. Finally, we characterize the resulting scaling behavior and the transition toward linear behavior, providing a theoretical rationale for previously reported empirical findings.

\paragraph{Kinematics and Geometric Configuration}
We adopt a global reference frame $(\mathbf{e}_{x}^\circ,\, \mathbf{e}_{y}^\circ)$ with the symmetry axis of the configuration aligned with $\mathbf{e}_{y}^\circ$. (For convenience, an auxiliary/local frame, $(\mathbf{e}_{x},\, \mathbf{e}_{y})$, will be introduced below.) Equal \textit{dead} tensions (i.e., prescribed constant forces applied independently of displacement) act at both ends of the elastica. At the rod extremity $C_2$, the tension vector is
\begin{equation}
\label{eq:vector_tension}
\mathbf{T}=T \cos\left(\frac{\phi}{2}\right)\, \mathbf{e}_{x}^\circ +  T \sin\left(\frac{\phi}{2}\right) \mathbf{e}_{y}^\circ,
\end{equation}
where $T=\lvert\mathbf{T}\lvert$ and the capstan angle $\phi$ defines the imposed direction of $\mathbf{T}$.
The tension at the other end, $C_1$, is applied symmetrically about the $\mathbf{e}_{y}^\circ$ axis. Since we are interested in the sliding force $F_0$ obtained during translation of the cylinder along its axis, we neglect tangential friction along the contact arc, focusing exclusively on normal contact forces. It then follows that the tangential component of the contact forces between the central cylinder and the rod are oriented along $\mathbf{e}_{z}^\circ$, and there is no tangential component of contact force in the $(\mathbf{e}_{x}^\circ,\mathbf{e}_{y}^\circ)$ plane. The ultimate goal of our model is to compute $R_{\text{tot}}$, which is the integral on the contact zone of the normal force exerted by the rod on the disk. The sliding threshold for the vertical pulling force $F$ (along $\mathbf{e}_{z}^\circ$) then follows from Amontons-Coulomb friction law: $F_0=\mu \, R_{\text{tot}}$, where $\mu$ is the friction coefficient. This linear relationship between $R_{\text{tot}}$ and $F_0$ will be validated through FEM simulations in Section \ref{sec:elastica_validation_results} (Fig.~\ref{fig:results_capstan_V2}a). The model requires that the dimensionless tying tension $\tilde{T}=T (R+r)^2 / (EI)$ be sufficiently large, so that continuous contact is established along the disk surface.

The continuous contact region between the elastica and the disk extends from point $B_1$ (at arclength $s_{B_1}$) to $B_2$ (at $s_{B_2}$), spanning a wrapping angle of $2\varphi^{*}$ centered at point $A$. The two symmetric contact-free regions extend from the contact boundaries ($B_1$ and $B_2$) to points where $\mathbf{T}$ is applied ($C_1$ and $C_2$), corresponding to arc-length intervals $s\in[s_{C_1}, s_{B_1}]$ and $s\in[s_{B_2}, s_{C_2}]$. For convenience, we introduce an auxiliary (local) frame $(\mathbf{e}_{x},\, \mathbf{e}_{y})$ obtained by rotating $(\mathbf{e}_{x}^\circ,\, \mathbf{e}_{y}^\circ)$ counterclockwise by $\varphi^{*}$, placing point $B_2$  on the $\mathbf{e}_{y}$ axis. In this rotated frame, the applied tension makes an angle $\beta = \phi/2 - \varphi^{*}$ with respect to $\mathbf{e}_{x}$, so that $\mathbf{T}=T\cos \beta\, \mathbf{e}_{x} + T \sin \beta\, \mathbf{e}_{y}$. This definition captures the physical evolution of the system as the tying tension $T$ increases and the loop tightens. The contact arc grows continuously from the low-tension limit, characterized by point contact ($\beta \to \phi/2$ and $\varphi^{*} \to 0$), to the high-tension limit where the contact is fully extended ($\varphi^{*} \to \phi/2$ and $\beta \to 0$).

To determine the relationship between $R_{\text{tot}}$ and $T$, we develop a solution based on the equilibrium equations for the planar elastica and compute the internal forces and moments along the continuous-contact region. In this region, the disk exerts a distributed normal pressure $p(s)$, directed radially outward and acting perpendicular to the rod centerline. At the two contact boundaries, localized normal point forces of magnitude $f$ arise. These contact forces, together with the symmetry of the configuration, allow us to match the solution across the boundary of the contact region.
We note that both our FEM simulations and previous studies \citet{grandgeorge_elastic_2022,grandgeorge_mechanics_2021} reveal a localized change in the normal tractions at these boundaries, and we idealize them as point forces in our model. Point forces at the boundaries of continuous-contact regions are a well-established concept in theories for elastic rods in contact \cite{coleman+swigon:2000,heijden+al:2001,grandgeorge_elastic_2022,grandgeorge_mechanics_2021}.
Before presenting the equilibrium equations, we note that similar computations have been performed by \citet{Singh2022} with unequal loads applied at the two ends and friction present in the $(\mathbf{e}_{x},\, \mathbf{e}_{y})$ plane. In contrast, our goal here is to arrive at a formula expressing $R_{\text{tot}}$ in terms of $T$.

We now proceed to the governing equilibrium equations for the planar elastica, written in the auxiliary frame 
$(\mathbf{e}_{x},\, \mathbf{e}_{y})$:
\begin{subequations}
\begin{align}
x'(s)  &= \cos \theta, &
y'(s)  &=\sin \theta,  &
EI \theta'(s)  & =m,  \label{eq:E1} \\
n_x'(s)+f_x^{\text{ext}}  &=0,  &
n_y'(s)+f_y^{\text{ext}}  &=0,  &
m'(s)  &= n_x \sin \theta - n_y \cos \theta, \label{eq:E2}
\end{align}
\end{subequations}
where $s$ is the arclength along the rod, with origin at point $A$ ($s_A=0$), $\theta(s)$ is the inclination angle that the tangent $\mathbf{t}(s)=(\cos \theta, \sin \theta)$ makes with $\mathbf{e}_{x}$, $(x(s),\,y(s))$ are the local coordinates of the centerline, $m(s)$ is the internal bending moment, and $(n_x(s),\,n_y(s))$ is the internal force. The prime notation $'$ denotes differentiation with respect to arclength $s$. By symmetry, we focus on the half-problem $s\in [0,s_{C_2}]$.
The external force $(f_x^{\text{ext}},\,f_y^{\text{ext}})$ accounts for contact pressure and the point forces at the boundaries:
\begin{subequations}
\begin{align}
f_x^{\text{ext}}(s)  &= p(s) \, \sin \theta(s) &
f_y^{\text{ext}}(s)  &= -p(s) \cos \theta(s)  &
\text{for} \, s & \in [0,s_{B_2}) \\
f_x^{\text{ext}}(s_{B_2})  &= 0 &
f_y^{\text{ext}}(s_{B_2})  &=-f \, \delta(s-s_{B_2})  &
\text{for} \, s & = s_{B_2} \\
f_x^{\text{ext}}(s)  &= 0 &
f_y^{\text{ext}}(s)  &= 0 &
\text{for} \, s & \in (s_{B_2},s_{C_2}).
\end{align}
\end{subequations}

We solve the equilibrium equations in the continuous-contact region $(A,B_2)$.
The kinematics of contact requires $x(s)=(R+r) \sin \theta(s)$, $y(s)=-(R+r) \cos \theta(s)$, and $\theta(s) = (s-s_{B_2})/(R+r)$ with our choice of orientation of the frame $(\mathbf{e}_{x},\, \mathbf{e}_{y})$, yielding $\theta'(s)=1/(R+r)$. The bending constitutive relation yields $m(s)=EI / (R+r)$, thus $m'(s)=0$, which implies $n_x(s) \sin \theta(s)-n_y(s) \cos \theta(s)=0$.
Computing $m''(s)$ from Eq.~(\ref{eq:E2}) gives 
\begin{equation}
\label{eq:m_prime_prime}
m''(s)=0=-p(s)+ n_t(s)/(R+r),
\end{equation}
where we introduced the internal tension $n_t(s)=n_x \cos \theta+n_y \sin \theta$ and now differentiate it to obtain
\begin{equation}
\label{eq:n_t_prime}
n_t'(s)=n_x'(s) \cos \theta + n_y'(s) \sin \theta + \theta'(s) \, [-n_x \sin \theta + n_y \cos \theta]=0.
\end{equation}
Eqs.~(\ref{eq:m_prime_prime}) and (\ref{eq:n_t_prime}) indicate that both $p$ and $n_t$ are independent of $s$, satisfying
\begin{equation}
p \, (R+r) = n_t. \label{eq:E4}
\end{equation}
Having established the uniform force distribution in the contact region, we next determine these constants by matching the boundary conditions.

Integrating the force equilibrium in Eq.~(\ref{eq:E2}) yields 
$n_x(s)=p (R+r) \cos \theta(s)+n_{x}^{*}$
and
$n_y(s)=p (R+r) \sin \theta(s)+n_{y}^{*}$, where $n_{x}^{*}$ and $n_{y}^{*}$ are integration constants.
Evaluating $n_t$ from these expressions gives 
$n_t=p \, (R+r) + n_{x}^{*} \cos \theta(s) + n_{y}^{*} \sin \theta(s)$.
For Eq.~(\ref{eq:E4}) to hold for all $s \in [0,s_{B_2})$, we must have $n_{x}^{*}=0=n_{y}^{*}$, which determines the internal forces uniquely. Just before the contact boundary, we obtain
\begin{equation}
\label{eq:E55}
n_x(s_{B_2}^-)=(R+r) \, p \cos \theta(s_{B_2})=(R+r) \, p
\: , \quad 
n_y(s_{B_2}^-)=(R+r) \, p \sin \theta(s_{B_2})=0.
\end{equation}
At the contact boundary $s=s_{B_2}$, the point force $f$ induces a jump in the internal force:
\begin{equation}
n_x(s_{B_2}^+) - n_x(s_{B_2}^-) + 0 = 0 \: , \quad 
n_y(s_{B_2}^+) - n_y(s_{B_2}^-) - f = 0. \label{eq:E5}
\end{equation}
These boundary conditions allow us to connect the solution in the contact region with the solution in the contact-free region.

In the contact-free region $s \in (s_{B_2},s_{C_2}]$, the internal force vector is uniform. Using $\theta(s_{C_2})=\beta$, the force boundary condition at $s_{C_2}$ yields
\begin{equation}
n_x = T \cos \beta \quad \text{ and }\quad
n_y = T \sin \beta. \label{eq:E6}
\end{equation}
Combining Eqs.~(\ref{eq:E55}), (\ref{eq:E5}) and (\ref{eq:E6}) gives
\begin{equation}
p = \frac{T}{R+r} \cos \beta \: \text{ and }\: 
f = T \sin \beta \label{eq:E7}
\end{equation}
These relations are independent of the wrapping angle $\varphi^{*}$ of the contact region. For a given tension $T$, Eqs.~(\ref{eq:E7}) provides two equations for three unknowns ($p$, $f$, and $\beta$), requiring an additional equation for closure.

\paragraph{Hamiltonian Conservation and Analytical Solution}
In principle, the required closure equation could be obtained from the moment equilibrium in Eq.~(\ref{eq:E2}), between $s=s_{B_2}$, where $m(s_{B_2})=EI/(R+r)$, and $s=s_{C_2}$, where $m(s_{C_2})=0$. However, since the solution $m(s)$ involves elliptic integrals, we use an alternative approach based on the conserved quantity
\begin{equation}
H=\frac12 \, \frac{m^2(s)}{EI} + n_t(s),
\end{equation}
which represents the Hamiltonian of the elastica and remains constant along the entire rod when tangential friction is neglected \cite{oreilly2017,Singh2022,Neukirch2025Noetherian}. We furthermore consider the long-length limit of the wrapping rod so that its ends are aligned with the applied tensions.
Evaluating $H$ at $s=s_{B_2}$ gives $H=(1/2) EI/(R+r)^2 + p (R+r)$, while evaluating it at $s=s_{C_2}$ yields $H=T$. Equating these expressions gives
\begin{equation}
T= \frac12 \frac{EI}{(R+r)^2} + p\, (R+r). \label{eq:E8}
\end{equation}
Combining the three relations in Eqs.~(\ref{eq:E7}) and (\ref{eq:E8}) enables us to write the key dimensionless loading relations
\begin{subequations}
\label{eq:non_dimensional_forces}
\begin{align}
\tilde{T} &= \frac{T\, (R+r)^2}{EI} = \frac12 \, \frac{1}{1-\cos \beta}, \label{eq:non_dimensional_T}
\\
\tilde{p} &= \frac{p\, (R+r)^3}{EI} = \frac12 \, \frac{\cos \beta}{1-\cos \beta}, \label{eq:non_dimensional_p}
\\
\tilde{f} &= \frac{f\, (R+r)^2}{EI} = \frac12 \, \frac{\sin \beta}{1-\cos \beta}. \label{eq:non_dimensional_f}
\end{align}
\end{subequations}
For both the pressure $p$ and the point force $f$ to remain positive, we require the angle $\beta \in  [0,\pi/2]$. As $\beta \to 0$, all three quantities $\tilde{T}$, $\tilde{p}$, and $\tilde{f}$ diverge, corresponding to the high-tension limit $T \gg EI/(R+r)^2$ in which the contact arc approaches its maximum extent around the cylinder.

Finally, we compute the total dimensionless normal force. By symmetry, the total force is twice that of the half-problem: 
\begin{equation}
\tilde{R}_\text{tot}=
\frac{R_\text{tot}\, (R+r)^2}{EI} =
2\left(\frac{1}{R+r}\int_{0}^{s_{B_2}} \tilde{p} ds +\tilde{f}\right) = \frac{\varphi^{*} \cos \beta + \sin \beta }{1-\cos \beta},  
\end{equation}
%
%where $\tilde{s}=s/(R+r)$.
%dont say twice : where the factor of 2 accounts for contributions from both halves of the symmetric configuration. 
Inverting Eqs.~(\ref{eq:non_dimensional_T}) yields 
$\beta=\arccos((2\tilde{T}-1)/(2\tilde{T}))$, bringing us to the central results of this analysis, the expressions for the total dimensionless normal force, $\tilde{R}_\text{tot}$, and the wrapping, $\varphi^{*}$, angle as functions of the magnitude of the applied tying tension $\tilde{T}$and the imposed capstan angle $\phi$:
\begin{subequations}\label{eq:planar_elastica_solution}
\begin{align}
\tilde{R}_\text{tot} &=(2\tilde{T}-1) \, \left[ 
\frac{\phi}{2} - \arccos \left( \frac{2\tilde{T}-1}{2\tilde{T}}\right)
\right] + \sqrt{4 \tilde{T} - 1}
\label{eq:planar_elastica_R_tot}, \quad \text{and} \\
\varphi^{*}  &= \frac{\phi}{2} - \arccos\left( \frac{2\tilde{T}-1}{2\tilde{T}} \right).
\label{eq:planar_elastica_varphi}
\end{align}    
\end{subequations}

\paragraph{Critical Contact Regime and Contact Evolution}
Through Amontons-Coulomb friction law, $F_0=\mu \, R_{\text{tot}}$, Eq.~(\ref{eq:planar_elastica_R_tot}) provides an explicit prediction for the sliding strength $\tilde{F}_0$ as a function of $\tilde{T}$ and $\phi$. (In Section~\ref{sec:elastica_validation_results}, we test this prediction against experiments, and both FEM and DER simulations.) In addition to the positivity conditions on $p$ and $f$, continuous contact requires $\varphi^{*}>0$, where $\varphi^{*}=\phi/2-\beta$, which imposes $\beta < \phi/2$. This condition is always satisfied for $\phi>\pi$. However, for $\phi<\pi$, a critical angle $\beta$ exists above which continuous contact cannot be sustained.
Consequently, a minimum tying tension is required for continuous contact:
\begin{subequations}\label{eq:planar_elastica_limit_tyingtension}
\begin{align}
\tilde{T} &\geq \frac{1}{2} \, \frac{1}{1-\cos (\phi/2)} 
\quad \text{if } \phi < \pi & 
\label{eq:planar_elastica_limit_tyingtension1} \\
\tilde{T} &\geq \frac{1}{2} 
\quad \phantom {\, \frac{1}{1-\cos (\phi/2)}}\text{if } \phi \geq \pi. &
\label{eq:planar_elastica_limit_tyingtension2}
\end{align}    
\end{subequations}
Notably, for $\phi\geq \pi$, the model predicts a universal minimum dimensionless tension of $\tilde{T}_{\text{min}} = 1/2$, independent of the capstan angle. For $\phi<\pi$ the wrapping angle satisfies $\varphi^{*} \to 0$ as $T$ approaches this minimum value, indicating that the contact region degenerates to a single-point. In contrast, for $\phi \geq \pi$, a finite contact length persists before detaching, even at $\tilde{T}=1/2$. These behaviors support the physical interpretation that tying tension, if insufficient relative to the rod’s bending stiffness, is unable to sustain continuous contact. This analytical prediction is fully consistent with the experimental observations in Section~\ref{sec:results_CH}, where the rod loses effective contact with the cylinder at low values of tension $\tilde{T}$.

\begin{figure}[h!]
    \centering
    \includegraphics[width=0.9\textwidth]{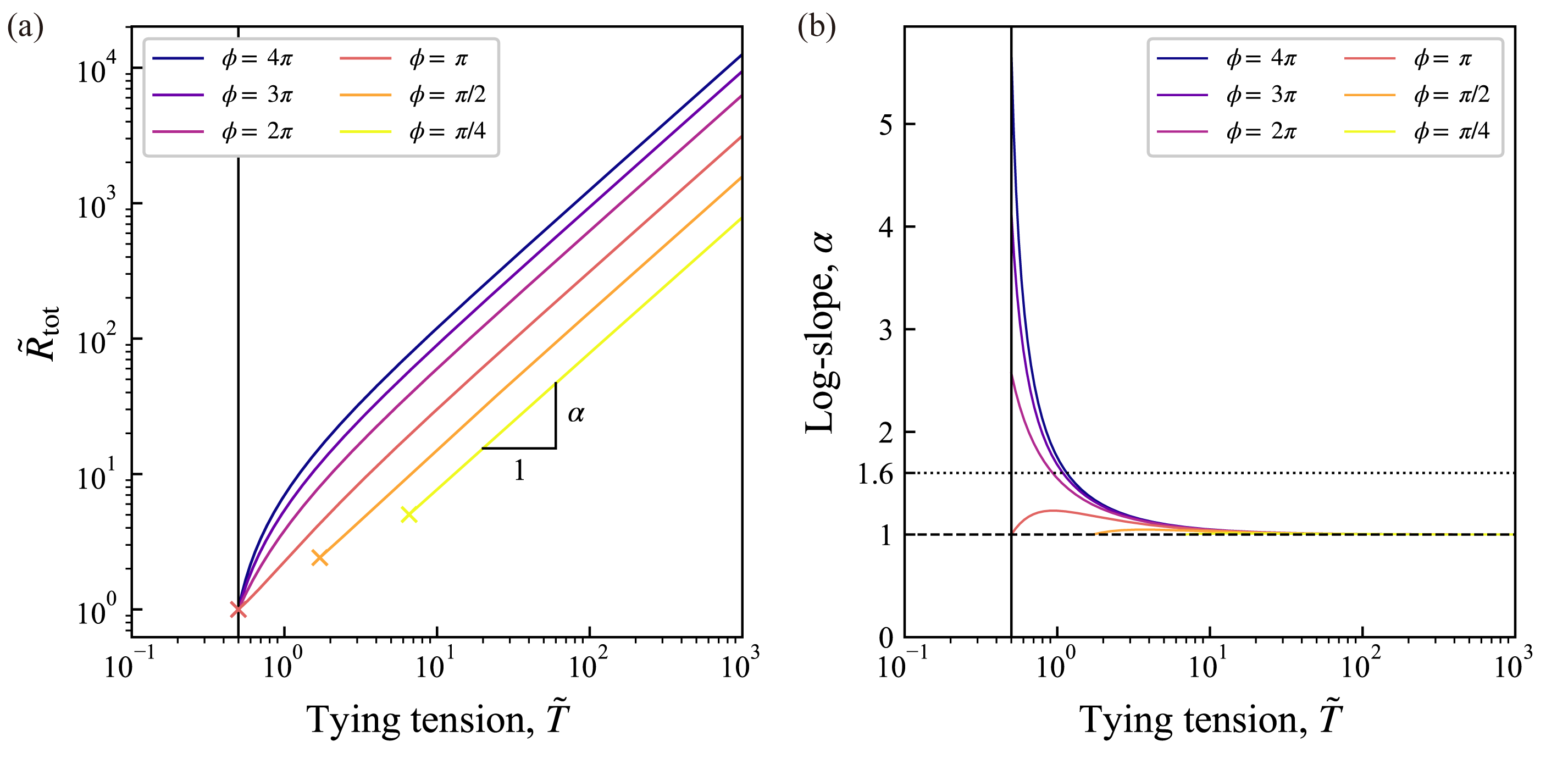}
    \caption{
    Analytical predictions from the planar elastica model, for several capstan angles $\phi\in[\pi/4,\, 4\pi]$. (a) Total dimensionless normal force $\tilde{R}_\text{tot}\left(\tilde{T}, \phi \right)$ obtained from the continuous contact solution of planar elastica, plotted over the feasible tying-tension range for each capstan angle. (b) Logarithmic derivative of the reaction with respect to the tying tension, $\alpha=\frac{\mathrm{d}(\log_{10} \tilde{R}_\text{tot})}{\mathrm{d} (\log_{10} \tilde{T})}$, quantifying the effective scaling exponent of the sliding strength. The dashed and dotted horizontal dotted correspond to $\alpha = 1$ and $1.6$.}
    \label{fig:results_analytical}
\end{figure}

\paragraph{Geometry-Driven Scaling Behavior and Transition to Linearity}
Figure~\ref{fig:results_analytical}a shows the analytical prediction of the total normal contact force $\tilde{R}_\text{tot}$ as a function of $\tilde{T}$ for capstan angles in the range $\phi \in [\pi/4, 4\pi]$. Cross markers denote the minimum admissible tension for continuous contact, defined by Eqs.~(\ref{eq:planar_elastica_limit_tyingtension}). As expected, $\tilde{R}_\text{tot}$ increases with the capstan angle, reflecting the greater wrapped arclength. At the onset of tightening, the initial slope becomes steeper with increasing $\phi$, indicating a more pronounced superlinear response before gradually transitioning toward linear growth at higher $\tilde{T}$.

To further analyze the functional dependence of $\tilde{R}_\text{tot}$ on $\tilde{T}$, we compute the slope $\alpha$ in the log-log scale:

\begin{equation}\label{eq:alpha}
\begin{aligned}
\alpha(\tilde{T})
&= \frac{\mathrm{d}(\log_{10} \tilde{R}_\text{tot})}{\mathrm{d}(\log_{10} \tilde{T})} = 1+\frac{\left[     \frac{\phi}{2} - \arccos\left( \frac{2\tilde{T}-1}{2\tilde{T}} \right)  \right]}{\tilde{R}_\text{tot}} 
= 1+\frac{\varphi^{*}}{\tilde{R}_\text{tot}}.
\end{aligned}
\end{equation}

Since $\varphi^{*}>0$ in the continuous contact regime, Eq.~(\ref{eq:alpha}) implies $\alpha>1$, confirming superlinear behavior. Figure~\ref{fig:results_analytical}b plots $\alpha(\tilde{T})$ for various capstan angles. The limiting values follow directly from Eqs.~(\ref{eq:planar_elastica_solution}) and (\ref{eq:alpha}):
%\begin{equation}
%\alpha(\tilde{T} \to \infty)=1,
%\end{equation}
%
\begin{subequations}
\label{eq:alpha_at_high_and_low}
\begin{alignat}{2}
\alpha\left(\tilde{T}=\frac{1}{2} \, \frac{1}{1-\cos (\phi/2)} \right) &=1 &\qquad& \text{for } \phi < \pi
\label{eq:alpha_at_0.5_low} \\
\alpha(\tilde{T}=1/2) &= 1-\pi/2+\phi/2 && \text{for } \phi \geq \pi 
\label{eq:alpha_at_0.5_high} \\
\alpha(\tilde{T} \to \infty) &=1 . &&
\end{alignat}    
\end{subequations}
These limiting results show that the degree of superlinearity near the onset of tightening increases monotonically with $\phi$ for $\phi \geq \pi$. Our analysis demonstrates that the nonlinearity observed is a consequence of the simultaneous evolution of contact loads ($\tilde{p}$ and $\tilde{f}$) and the wrapping angle ($\varphi^{*}$). As $\tilde{T}$ increases and the contact arc saturates ($\varphi^{*} \to \phi/2$, $\beta \to 0$), the system naturally transitions toward a linear regime ($\alpha \rightarrow  1$). 

This transition from nonlinear to linear behavior offers a rationale for the empirical power-law reported by \citet{johanns_strength_2023}, where a characteristic scaling exponent $\alpha \simeq 1.6$ and a subsequent shift to $\alpha = 1$ were observed for elastoplastic filaments. While our purely elastic theory predicts an increase in the initial slope with $\phi$, the experimentally observed nearly constant $\alpha$, independent of the number of loops $n$ (where $\phi= 2n\pi$), can be rationalized by noting that each loop in a surgical knot is plastically set into a discrete circular unit, rather than forming a single continuous helix. Our model thus suggests that each loop responds as an independent single-loop capstan ($\phi=2\pi$). A detailed comparison is discussed in \ref{sec:appendix_comparison_surgical_knots}.

%%%%%%%%%%%%%%%%%%%%%%%%%%%%%%%%%%%%%
\section{Validation of the elastica model and parametric study of capstan configurations} \label{sec:elastica_validation_results}
%%%%%%%%%%%%%%%%%%%%%%%%%%%%%%%%%%%%%

We now validate the analytical relationship between $\tilde{R}_\text{tot}$, $\tilde{T}$, and $\phi$ established in (Eq.~\ref{eq:planar_elastica_R_tot}) against experimental and numerical results. To explore how the capstan angle governs the system's mechanics, we conduct a systematic parametric study across the range $\phi\in[\pi/4,\, 4\pi]$.
\begin{figure}[b!]
    \centering
    \includegraphics[width=0.9\textwidth]{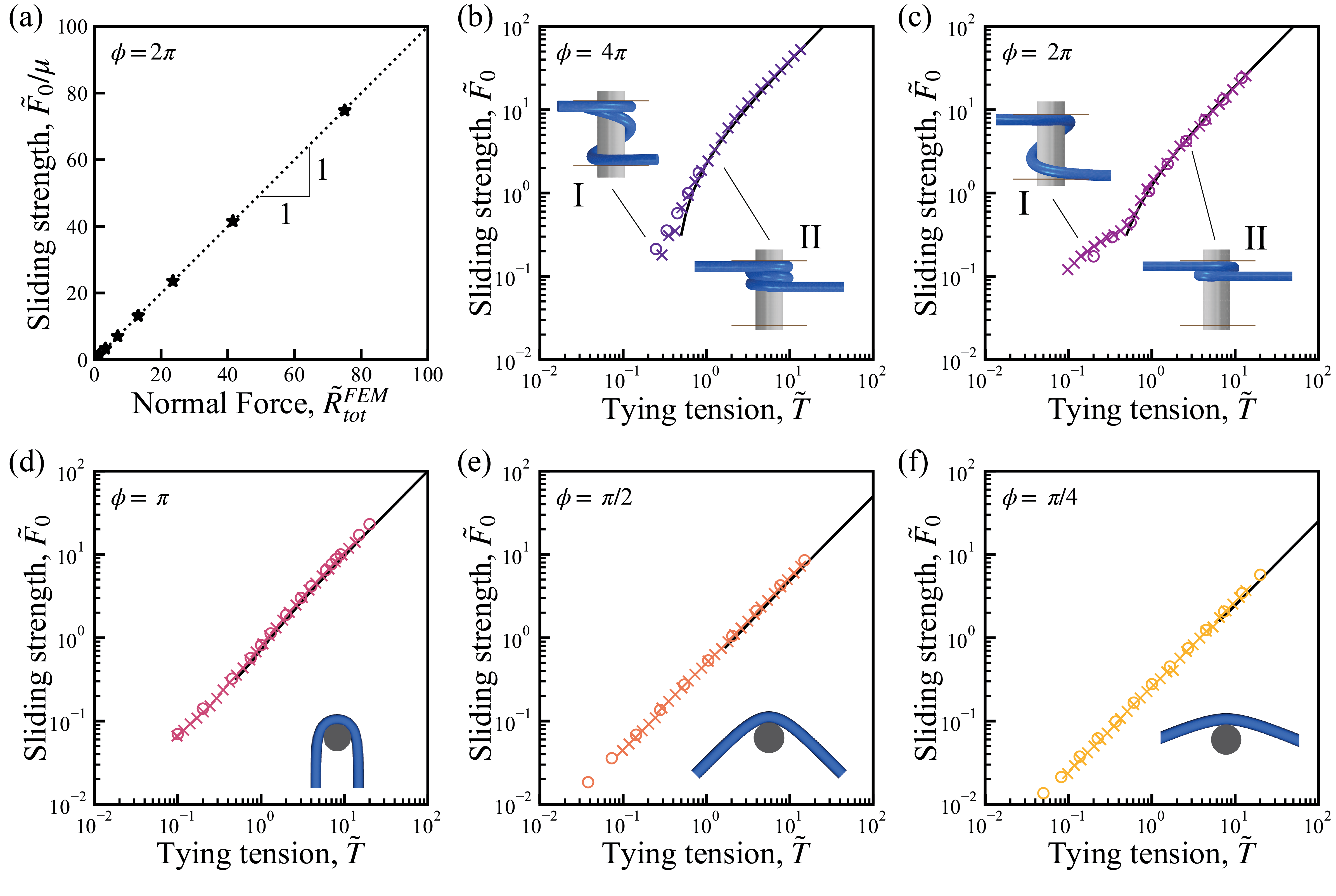}
    \caption{Numerical results for sliding strength for multiple capstans.
    (a) Normalized sliding strength divided by the friction coefficient, $\tilde{F}_0/\mu$, plotted against the total normal force obtained from FEM, $\tilde{R}^{FEM}_\text{tot}$.
    (b–f) Normalized sliding strength $\tilde{F}_0$ as a function of tying tension $\tilde{T}$ for capstan angles (b) $4\pi$, (c) $2\pi$, (d) $\pi$, (e) $\pi/2$, and (f) $\pi/4$. (circles): FEM, (crosses): DER. The solid lines denote predictions from the planar elastica model, $\tilde{F}_0=\mu\tilde{R}_{tot}$, where $\tilde{R}_{tot}$ is given in Eq.~(\ref{eq:planar_elastica_R_tot}).}
    \label{fig:results_capstan_V2}
\end{figure}

First, we  verify whether the sliding strength $\tilde{F}_0$ varies linearly with the normal contact force $\tilde{R}_\text{tot}$ transmitted at the rod-cylinder interface, consistent with the Amontons–Coulomb friction relation $\tilde{F}_0=\mu \tilde{R}_\text{tot}$, where $\mu$ denotes the friction coefficient. Figure~\ref{fig:results_capstan_V2}a shows the FEM results for the one-loop configuration ($\phi = 2\pi$), demonstrating a clear linear dependence between $\tilde{F}_0/\mu$ and normal force $\tilde{R}^{FEM}_\text{tot}$ obtained from FEM, for a value of friction coefficient between VPS (rod) and POM (cylinder) of $\mu=0.32$. This linearity confirms that the planar elastica model provides a reliable estimate of the normal force, thereby enabling direct comparison between theoretical predictions of $\tilde{R}_\text{tot}$ and the sliding forces $\tilde{F}_0$ measured in experiments and simulations.

Figure~\ref{fig:results_capstan_V2}b–f shows that the FEM and DER results agree closely with the analytical prediction in Eq.~(\ref{eq:planar_elastica_R_tot}) across capstan angles $\phi = \{ 4\pi,~2\pi,~\pi,~\pi/2,~\pi/4\}$. The panels include schematic views of each configuration; notably, (b) and (c) illustrate both the initial loose and final tightened states, while the others focus on the tightened geometry. The model predicts a continuous-contact regime bounded below by Eq.~(\ref{eq:planar_elastica_limit_tyingtension}); below this limit, no solution for $R_\text{tot}$ (and thus $\tilde{F}_0$) exists. For $\phi > \pi$, the response becomes markedly sublinear at low tensions ($\tilde{T} < 0.5$), leading to an abrupt change in slope, visible as a kink at $\tilde{T} = 0.5$ in the $\tilde{F}_0$–$\tilde{T}$ curves. As confirmed by FEM, this nonlinearity arises because configurations with one or more full loops relax from a planar coil into a three-dimensional helical shape at low tensions (see Insets II and I, respectively, in Fig.~\ref{fig:results_capstan_V2}(b,c)), causing a loss of continuous contact. By contrast, for $\phi \le \pi$ the configuration remains planar across the entire range of $\tilde{T}$, and no such transition is observed.

\begin{figure}[h!]
    \centering
    \includegraphics[width=0.9\textwidth]{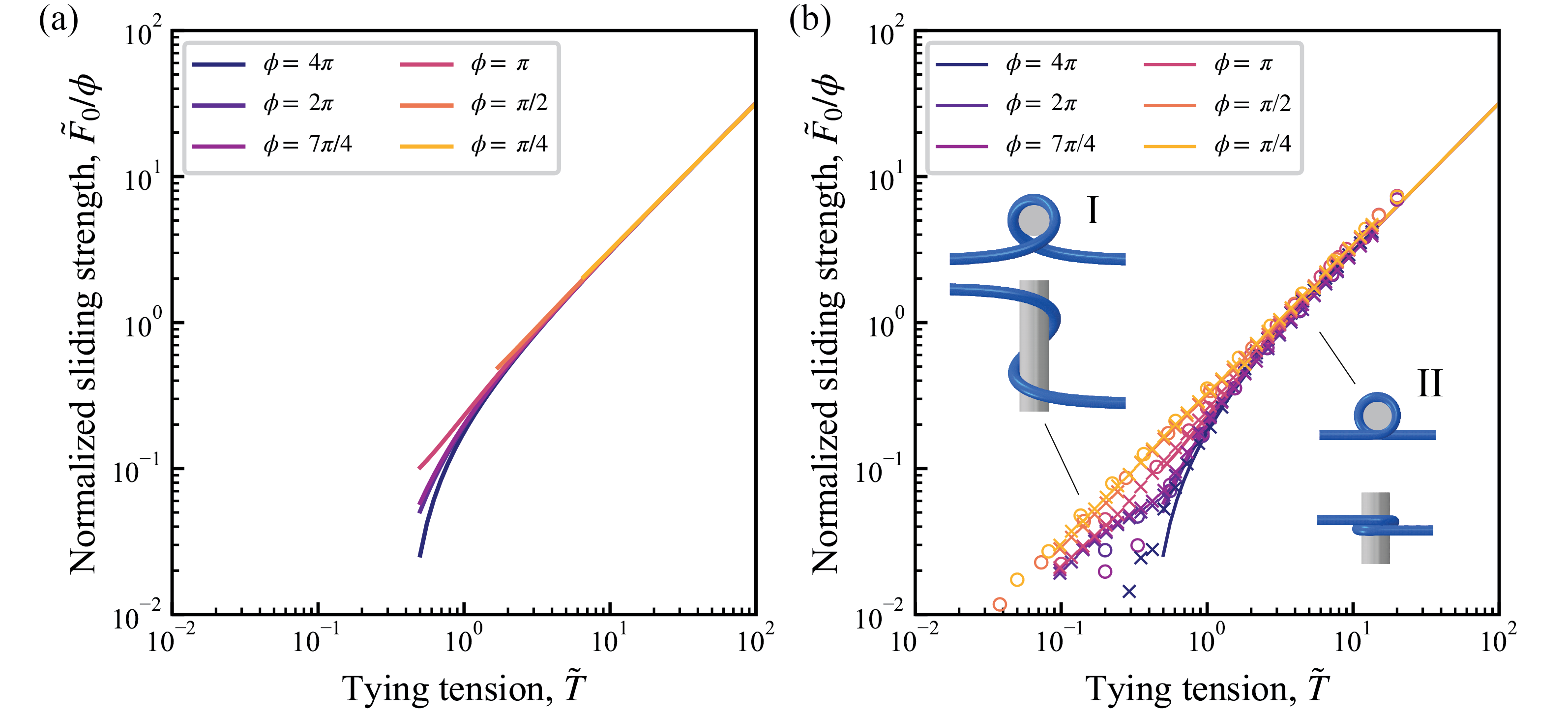}
    \caption{
    (a) Analytical prediction of normalized sliding strength, $\tilde{F}_0/\phi$, plotted against tying tension $\tilde{T}$ for various capstan angles $\phi = \{4\pi,~2\pi,~\pi,~\pi/2$, $\pi/4\}$.
    (b) Numerical validation via FEM and DER results superimposed on the analytical curve, demonstrating collapse across capstan angles. Insets provide representative FEM visualization of the rod configurations: a loose, helical geometry at low tension (left) and a tight, two-dimensional configuration at high tension (right). The FEM and DER data are identical to those presented in Fig.~\ref{fig:results_capstan_V2}.}
    \label{fig:results_capstan_collapsed}
\end{figure}

To further illustrate this transition from a planar coil to a 3d helix, Fig.~\ref{fig:results_capstan_collapsed}a presents the analytical predictions plotted as $\tilde{F}_0/\phi$ versus $\tilde{T}$. At high tying tensions where bending stiffness is negligible, all curves collapse onto a single linear master curve: in the limit $\tilde{T} \to \infty$, Eq.~(\ref{eq:planar_elastica_R_tot}) yields $\tilde{R}_\text{tot}/\phi \to \tilde{T}$. Figure~\ref{fig:results_capstan_collapsed}b confirms this collapse when the FEM and DER results are superimposed on the analytical prediction. This linear collapse further implies that the high-tension regime can be used to experimentally infer the friction coefficient $\mu$ by measuring the sliding strength and extracting the slope from a linear-linear plot, according to $\mu=\tilde{F}_0/(\phi\tilde{T})$. Consequently, the robust data collapse across varying capstan angles confirms the accuracy of the 2D planar elastica model, especially for $\tilde{T} \geq 1$ as the system moves beyond the bending-dominated regime into a tension-governed regime.

\section{Characterization of contact} 
\label{sec:contact} 

The planar elastica model provides key insight into the mechanics governing sliding strength in capstan configurations. We now extend this understanding to more complex settings, as in the clove hitch, which involves fully 3D geometry and cross-sectional deformation. To this end, we analyze the contact interactions in knotted structures by comparing FEM results across three representative cases: the $\pi$ and $2\pi$ capstan configurations and the clove hitch.

\begin{figure}[b!]
    \centering
    \includegraphics[width=0.9\textwidth]{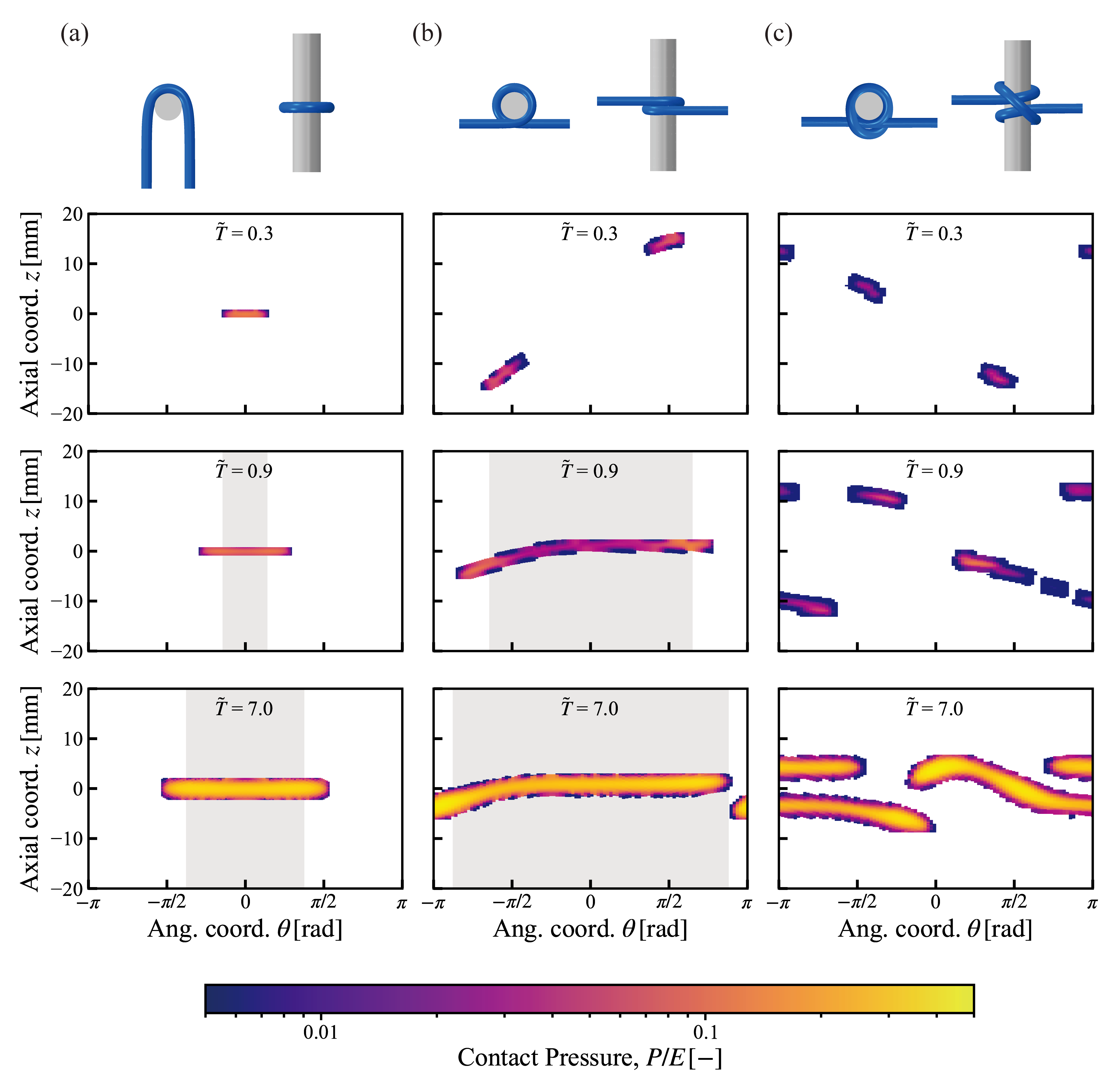}
    \caption{Graphical illustration and numerical analysis of the contact interface for various rod configurations (a) $\phi = \pi$, (b) $\phi = 2\pi$ and (c) the clove hitch. The 2D maps display the non-dimensionalized contact pressure $P/E$ as a function of the cylinder's angular coordinate $\theta$ and the axial coordinate, $z$. Results are shown for three representative tension levels, $\tilde{T}=\{0.3,~0.9,~7.0\}$. The gray shaded regions in (a) and (b) denote the theoretical contact zones predicted by the planar elastica model.}
    \label{fig:Contact_evolution}
\end{figure}

Figure~\ref{fig:Contact_evolution}
presents snapshots of the contact pressure distribution on the vertical cylinder for (a) the $\pi$, (b) the $2\pi$, and (c) the clove hitch configurations, at three applied tensions, $\tilde{T}=0.3$, $0.9$, and $7.0$. For the tensions within the continuous contact regime ($\tilde{T}=\{0.9,~7.0\} \ge 0.5$), the FEM results are superimposed with analytical predictions from the planar elastica model, represented by the shaded grey regions, which indicate the theoretical contact arc $[-\varphi^{*}, \varphi^{*}]$, predicted in Eq.~(\ref{eq:planar_elastica_varphi}). 

In the $\pi$ configuration (first column of Fig.~\ref{fig:Contact_evolution}), the FEM solution begins with a single-point contact that gradually develops into a line contact as the tying tension increases. While the planar elastica predicts continuous-contact solutions only for $\tilde{T} \ge 0.5$, the FEM results exhibit finite contact area even for $\tilde{T} < 0.5$. This initial discrepancy is primarily due to the rod's cross-sectional deformation, which locally smears the contact. The effect becomes more pronounced at higher tensions, where increased contact pressure leads to wider axial ($z$-direction) contact and a contact arc that exceeds theoretical predictions. Despite these local deviations, the FEM results follow the predictions provided by the elastica model in a semi-qualitative manner, once in the continuous-contact regime ($\tilde{T} \geq 0.5$). In this regime, both the contact pressure and the arclength grow monotonically with $\tilde{T}$, until the arc eventually saturates at the geometric limit, $\varphi^{*} = \phi/2=\pi/2$.

The full 3D FEM for the $2\pi$ configuration (center column of Fig.~\ref{fig:Contact_evolution}) exhibits more complex behavior. At at $\tilde{T}=0.3$, we observe the previously discussed helical configuration (Fig.~\ref{fig:results_capstan_V2}), which transitions to a planar shape for $\tilde{T} \geq 0.5$. The initial contact pressure distribution along the contact line is markedly non-uniform, displaying a pronounced maximum near the entry point of the wrap, consistent with the force jump $f$ at the contact boundaries, assumed in the model presented in Section~\ref{sec:elastica_model}. With increasing tension, this distribution progressively flattens, as the contribution of the jump force diminishes ($\tilde{f}/\tilde{p}=\tan\beta \to 0$ as $\beta \to 0$, see Eq.~(\ref{eq:non_dimensional_forces})), leading to an increasingly uniform pressure profile.

The clove hitch (right column of Fig.~\ref{fig:Contact_evolution}) exhibits a contact evolution that is qualitatively similar to the capstan configurations with $\phi>\pi$, despite its more complex three-dimensional geometry and self-contact. At low tension ($\tilde{T}<0.5$), the rod adopts a loose shape with discrete contact points, which merge into two continuous contact loops as $\tilde{T}$ increases. This progression aligns with the same tension-driven transition observed in simple capstan systems and demonstrates that the underlying mechanism persists even in geometries beyond the idealized capstan.

\section{Discussion and Conclusions}
\label{sec:conclusions}

In this study, we systematically investigated the sliding strength ($\tilde{F}_0$) of a thin filament wrapped around a cylinder sliding perpendicularly to the filament axis. Motivated by the nonlinear sliding behavior originally identified in elasto-plastic surgical filaments by \citet{johanns_strength_2023}, we clarified the physical origin of this response, specifically examining the necessity of material plasticity, the validity of power-law interpretations, and the underlying mechanics governing the interaction. To address these questions, we combined model experiments, numerical simulations (FEM and DER), and a new theoretical framework based on the planar elastica.

We first examined the clove hitch knot tied with a wide range of filament materials, ranging from elastic rods (elastomeric polymers, metals) to inelastic rope, and found the same superlinear scaling in all cases. This material-independence demonstrated that plasticity was not responsible for the observed scaling. At sufficiently high tying tensions ($\tilde{T}>10$), the response asymptotically converged to a linear regime, consistent with the expected scaling of normal forces under strong tightening. Experiments on the $2\pi$ capstan configuration exhibited the same transition, indicating that the nonlinear scaling was not specific to knot topology.

To verify and extend these experimental findings, we performed 3D FEM and DER simulations. The FEM captured the out-of-plane geometry and provided detailed information on contact interactions, while the DER neglected cross-sectional deformation. The fact that both numerical approaches reproduced the same experimentally observed scaling behavior evidences that local Hertzian contact deformations \cite{hertz1881contact} are not the cause of the observed nonlinearity. Notably, the sliding mechanics were found to be robustly captured by our framework, even for $R/r$ as small as unity (see \ref{sec:appendix_radius_ratio}). This macroscopic robustness was further justified by the evolution of the contact wrapping angle (see \ref{sec:appendix_contact_set_evolution}). For $\tilde{T} \ge 0.5$, the wrapping angle closely followed the trend of the planar elastica solution, despite some fluctuations at high tensions. These fluctuations in the wrapping arclength, primarily driven by 3D self-capsizing, are internally balanced by an increase in local contact pressures, ensuring that the macroscopic mechanics are governed by the primary interplay between bending and tension.

To rationalize these observations, we introduced an analytical model based on the planar elastica for a filament wrapped around a rigid cylinder. The solution predicted the emergence of a superlinear regime for the sliding strength as a function of applied tension for values of $\tilde{T} \ge 0.5$, where the configuration became effectively planar and continuous line contact was established. Within this validity range, analytical predictions agreed with both experimental and numerical results across multiple capstan configurations, including cases with $\phi > \pi$ that remained slightly non-planar even at high tensions. For $\phi > \pi$, a distinct strengthening of the sliding response appeared at $\tilde{T}=0.5$ in experimental results, marking the transition from helical 3D configurations to a planar, coiled state.

Together, these results elucidated the physical mechanism underlying the observed nonlinear scaling. The superlinear regime originated from the coupled evolution of the normal force and the contact arclength as the filament is tightened around the cylinder. The planar elastica model successfully captured this interplay, demonstrating that the transition to linear scaling occurs once the contact arclength saturates. Finally, we validated our model against the experimental results for surgical knots reported by \citet{johanns_strength_2023}. While the empirical scaling exponent $\alpha \simeq 1.6$ remained nearly constant regardless of the number of loops, our model accounted for this invariance by treating the knot as a superposition of independent $2\pi$ wraps. This approach was justified by the fact that each hitch in a surgical knot is plastically set into a planar circular geometry, effectively decoupling its mechanical response from adjacent loops (see \ref{sec:appendix_comparison_surgical_knots} for details).

We have demonstrated that the previously reported power-law behavior arises as a predictable consequence of the interplay between elastic bending and evolving contact geometry during tightening. The consistent agreement among theory, simulations, and experiments provides a solid foundation for understanding the frictional transmission in flexible filaments and offers practical guidelines for applications, such as surgical knotting, where controlling knot security requires precise, quantitative knowledge of how the tying tension governs mechanical strength.

\section*{Acknowledgments}

We are grateful to John Maddocks for enlightening discussions and to Ubamanyu (Uba) Kanthasamy for assistance with preliminary aspects of setting up the FEM simulations.

\appendix

\section{Sensitivity of sliding strength to the radius ratio $R/r$}
\label{sec:appendix_radius_ratio}

\begin{figure}[h!]
    \centering
    \includegraphics[width=0.8\textwidth]{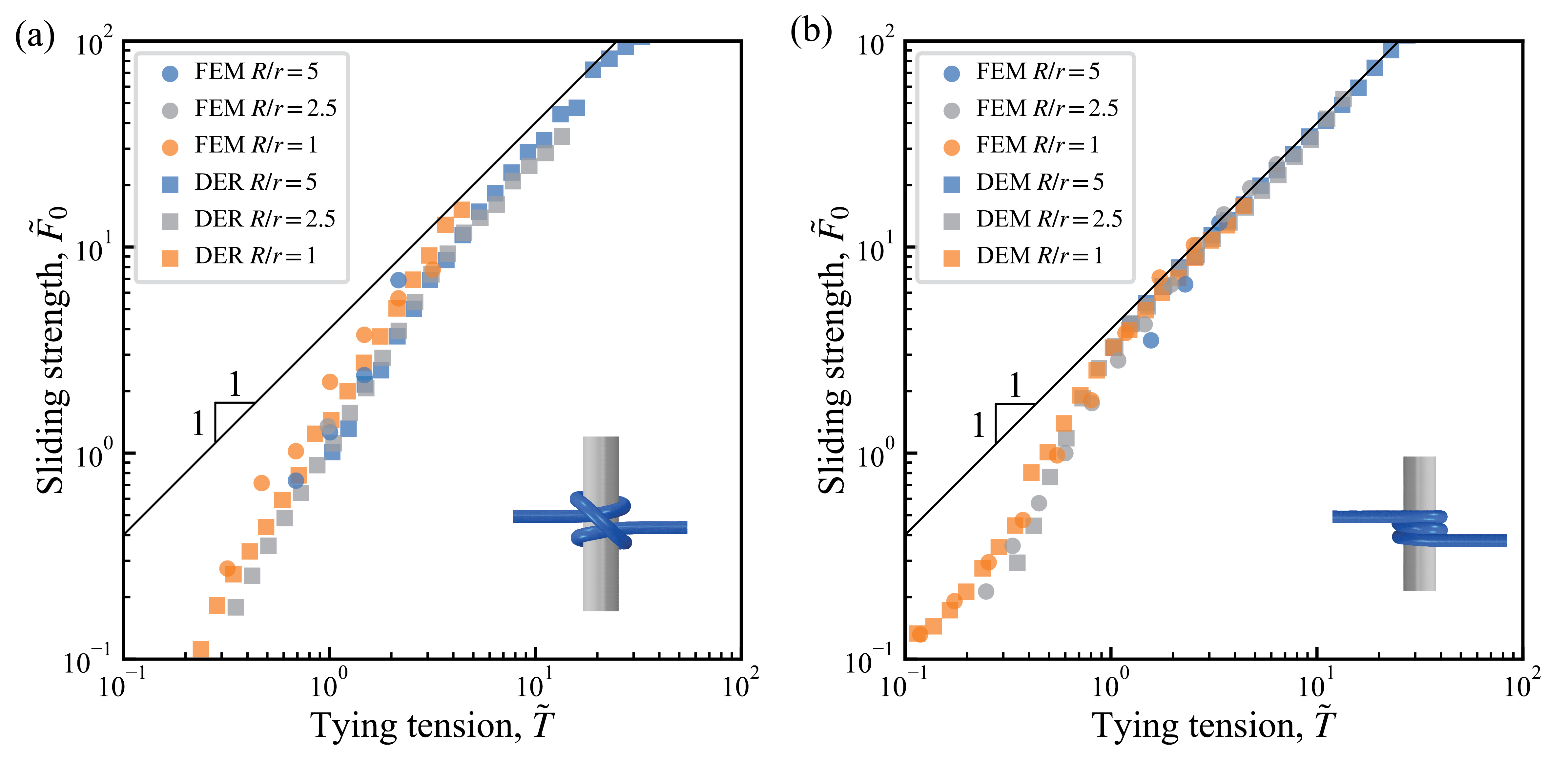}
    \caption{Sensitivity of sliding strength to the radius ratio $R/r$. Dimensionless sliding strength $\tilde{F}_0$ versus tying tension $\tilde{T}$ for (a) the clove hitch and (b) the $\phi=4\pi$ capstan, for $R/r \in \{1.0, 2.5, 5.0\}$, comparing DER and FEM results.}
    \label{fig:Knotstrength_Rr}
\end{figure}

We investigate the sensitivity of the sliding strength $\tilde{F}_0$ to the radius ratio $R/r \in \{1, 2.5, 5\}$ to assess whether cross-sectional deformation or localized distortion affects the sliding mechanics. Reducing $R/r$ increases contact curvature and localized pressure, providing a direct test for the limits of our centerline-based elastic theory. As shown in Fig.~\ref{fig:Knotstrength_Rr}a, the data for the dimensionless sliding strength collapse across all tested geometric ratios and numerical methods, for both the clove hitch configurations and the $\phi=4\pi$ capstan. The close agreement between the DER (rigid cross-sections) and the FEM (full 3D deformation) confirms that the global force balance is primarily governed by the interplay of bending and tension, while local cross-sectional distortion remains a secondary effect.

\section{Contact set evolution in the $\phi=4\pi$ configuration}
\label{sec:appendix_contact_set_evolution}

\begin{figure}[h!]
    \centering
    \includegraphics[width=0.75\textwidth]{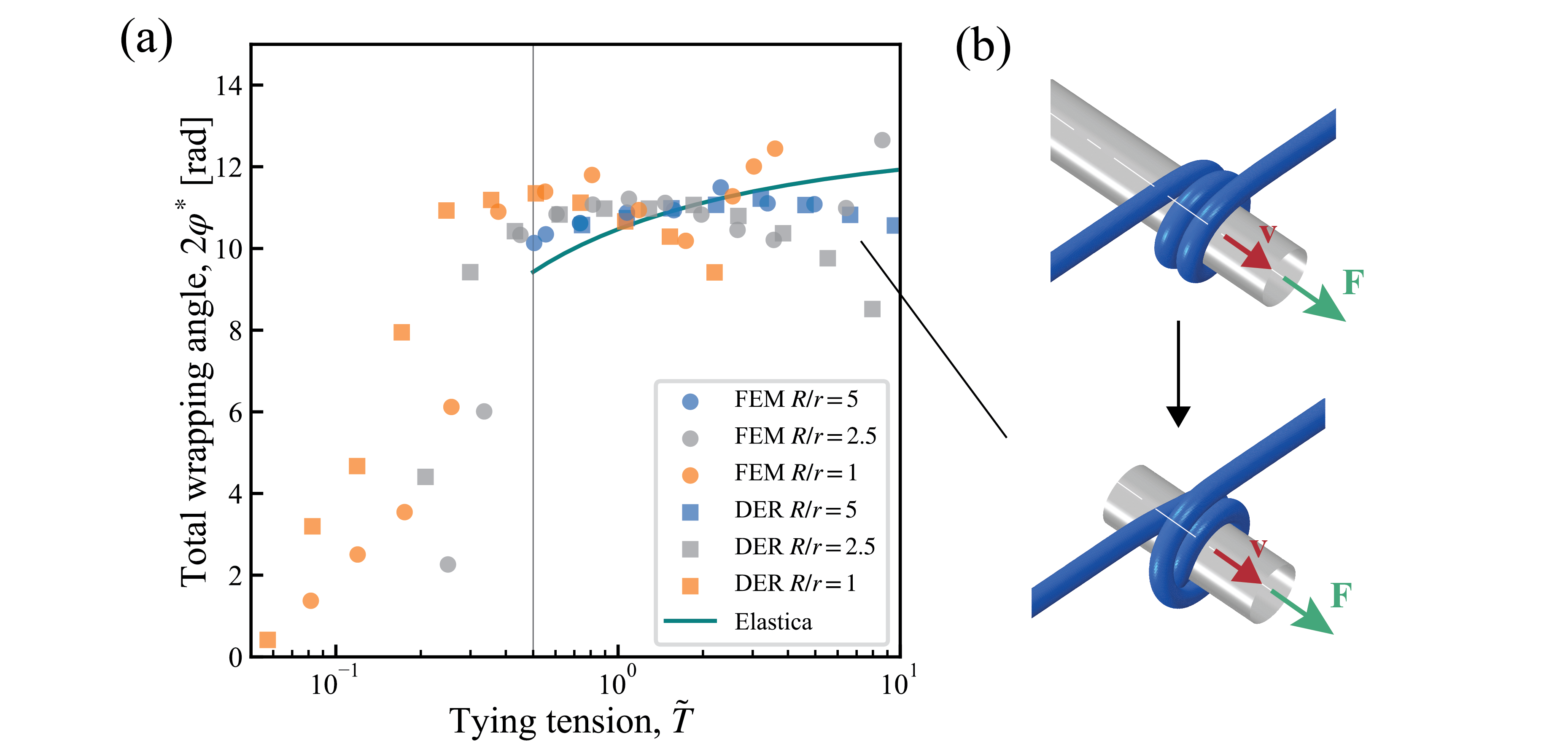}
    \caption{(a) Evolution of the total wrapping angle $2\varphi^*$ \textit{vs.} tying tension $\tilde{T}$ for the $\phi=4\pi$ capstan. The solid line is the contact angle predicted by the planar elastica model. The vertical gray line marks the theoretical onset of continuous contact at $\tilde{T}=0.5$. (b) FEM visualizations of the configuration at $\phi=4\pi$ and $\tilde{T}=5$, contrasting the initial tightened state prior to cylinder translation (top) with the self-capsizing state that develops during sliding (right).}
    \label{fig:wrapping-angle_Rr}
\end{figure}

With the robustness of the sliding strength to geometric ratio established in~\ref{sec:appendix_radius_ratio}, we now examine the underlying contact mechanics by analyzing the evolution of the contact set for the $\phi=4\pi$ configuration. In Fig.~\ref{fig:wrapping-angle_Rr}, we plot the projected total wrapping angle $2\varphi^*$, revealing two distinct kinematic regimes. For $\tilde{T} \ge 0.5$, the wrapping angle shows relatively good agreement with the planar elastica model, asymptotically approaching the geometric limit ($2\varphi^{*} \to 4\pi$ as $\tilde{T} \to \infty$) as continuous line contact is established; this trend remains consistent even as $R/r$ decreases from 5 to 1. For $\tilde{T} < 0.5$, the wrapping angle decreases sharply as the system enters a loose, 3D configuration that eventually transitions toward discrete point contacts. While the numerical data at higher tensions show noticeable fluctuations from the theoretical monotonic saturation, the overall trend is preserved. This discrepancy is attributed to 3D self-capsizing (Fig.~\ref{fig:wrapping-angle_Rr}b), where portions of the filament overlap or penetrate beneath adjacent loops during sliding, reducing the effective contact arclength. However, this loss of contact is compensated by an accompanying increase in local contact pressure at the overlapping segments, preserving the global force balance and further supporting the applicability of our macroscopic framework despite small-scale 3D complexities.

\section{Comparison with surgical knots}
\label{sec:appendix_comparison_surgical_knots}

\begin{figure}[h!]
    \centering
    \includegraphics[width=0.9\textwidth]{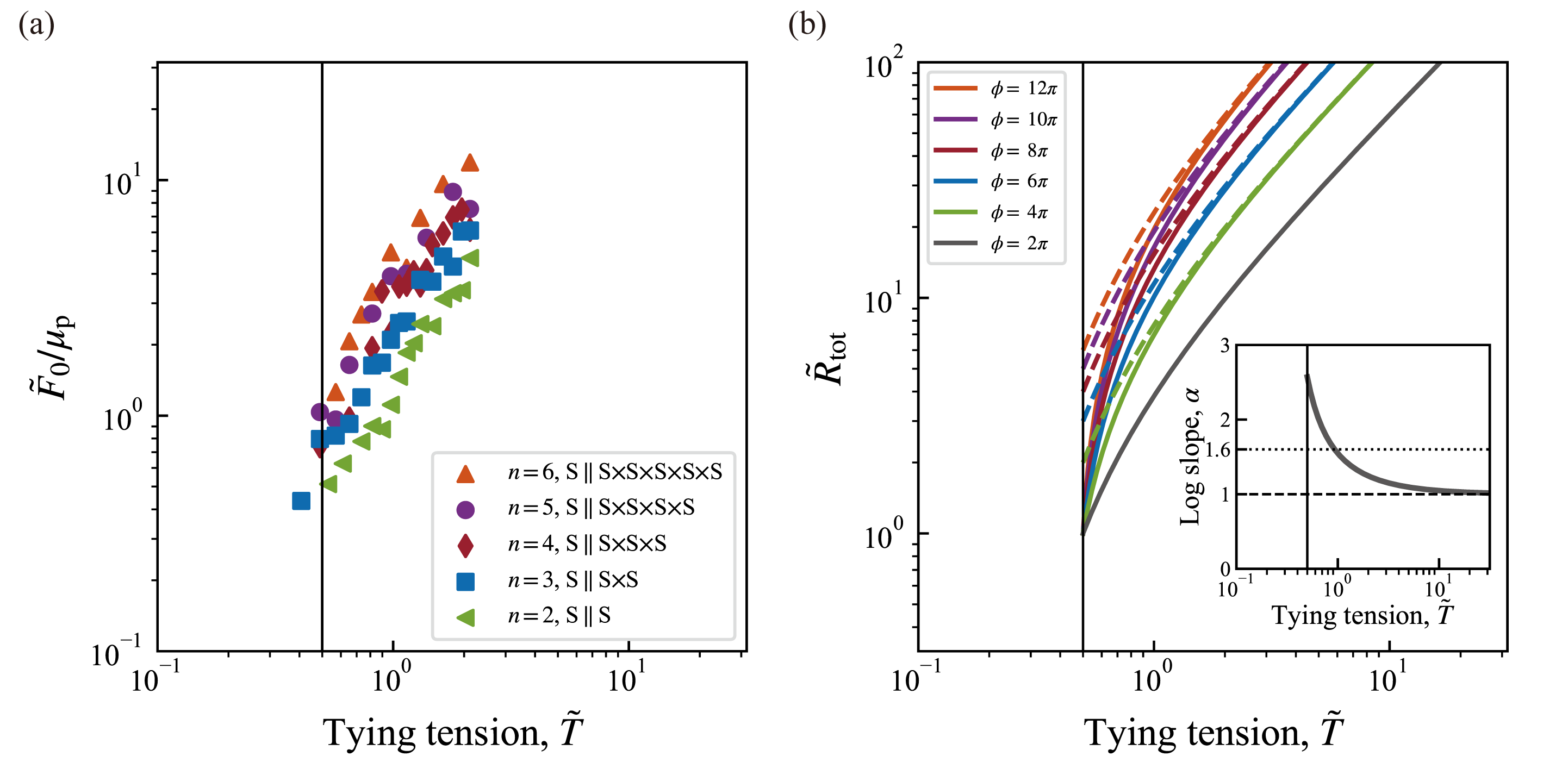}
    \caption{(a) Non-dimensional sliding strength $\tilde{F}_{0}/\mu_\text{p}$ versus tying tension $\tilde{T}$ for surgical knots with $n \in \{2,~3,~4,~5,~6\}$ loops, reevaluated from \cite{johanns_strength_2023}. The vertical black line marks the theoretical onset of continuous contact at $\tilde{T}=0.5$. (b) Total normal force $\tilde{R}_\text{tot}$ for $\phi = 2n\pi$ (solid lines) versus the superposition model $n\tilde{R}_\text{tot}(\phi = 2\pi)$ (dashed lines). Inset: log-slope $\alpha$ for a $2\pi$ wrap.}
    \label{fig:Surgical_Knots}
\end{figure}

Returning to the original motivation of the study, we revisit the nonlinear behavior observed in surgical knots reported by \citet{johanns_strength_2023}. To align the experimental data with our framework, we recast the results for different number of loops using our non-dimensional tension $\tilde{T}=T(R+r)^2/(EI)$ and compare them with the non-dimensional sliding strength, defined as $\tilde{F}_{0}/\mu_\text{p}=(F_{0}/\mu_\text{p})(R+r)^2/(EI)$ with $\mu_\text{p}=0.2$ denoting the friction coefficient for polypropylene suturing filaments. Here, $E=2\,$GPa represents the average modulus (up to 0.1 strain) derived from the initial loading cycle of the filament~\cite{johanns_strength_2023}. This non-dimensionalization reveals that the experimental tying tension lies within $0.5<\tilde{T}<2.5$, a range bounded by the low-tension contact limit ($\tilde{T}=0.5$ from Eq.~\ref{eq:planar_elastica_limit_tyingtension2}) and the ultimate fracture of the filament (Fig.~\ref{fig:Surgical_Knots}a). 

However, while the experiments confirm a superlinear relationship, the sensitivity of this superlinearity to the number of loops is less pronounced than predicted by our purely elastic theory. This weaker sensitivity is not simply due to experimental uncertainty but reflects the distinct structural role of plasticity: while superlinear scaling is an intrinsic feature of elastic rod-cylinder interaction and does not require plastic deformation, in surgical practice each hitch is plastically set into a circular geometry, effectively decoupling its mechanical response from adjacent loops and preventing unraveling during subsequent throws. This mechanical decoupling motivates a superposition approach where the total reaction force for an $n$-loop knot is modeled as $n\tilde{R}_\text{tot}(\phi = 2\pi)$ instead of a single continuous wrap $\tilde{R}_\text{tot}(\phi = 2n\pi)$. As illustrated in Fig.~\ref{fig:Surgical_Knots}b, this superposition (dashed lines) suggests a consistent superlinearity regardless of $n$. Notably, the power-law exponent $\alpha$ for a $2\pi$ wrap ranges from $1+\pi/2$ down to 1. At $\tilde{T} \approx 1$, our model yields $\alpha \approx 1.6$, in agreement with the previously reported empirical value of \citet{johanns_strength_2023}. Finally, we note that the predicted reaction forces systematically exceed the experimental values, which we attribute to the difference in boundary conditions: unlike the constant-tension loading assumed here, surgical knots undergo tension release and elastic spring-back after tying, reducing the effective tying tension at the moment of measurement.

%\noindent \textbf{Acknowledgments:} 
\bibliographystyle{elsarticle-num-names}
\bibliography{references}

@article{crassous_discrete_2023,
  title={Discrete-element-method model for frictional fibers},
  author={Crassous, J{\'e}r{\^o}me},
  journal={Physical Review E},
  volume={107},
  number={2},
  pages={025003},
  year={2023},
  publisher={APS}
}

@article{bergou.2008,
	Author = {Bergou, M. and Wardetzky, M. and Robinson, S. and Audoly, B. and Grinspun, E.},
	Journal = {ACM Transactions on Graphics},
	Number = {3},
	Pages = {1--12},
	Title = {Discrete elastic rods},
	Volume = {27},
	Year = {2008}
	}

@article{cundall.1979,
author = {Cundall, P. A. and Strack, O. D. L.},
title = {A discrete numerical model for granular assemblies},
journal = {Géotechnique},
volume = {29},
number = {1},
pages = {47-65},
year = {1979}
}

@book{Bigoni2012,
  author = {Davide Bigoni},
  publisher = {Cambridge University Press},
  title = {Nonlinear Solids Mechanics},
  year = {2012}
}

@article{tani2024,
  title = {How a Soft Rod Wraps around a Rotating Cylinder},
  author = {Marie Tani and Hirofumi Wada},
  journal = {Physical Review Letters},
  volume = {132},
  number = {5},
  pages = {058204},
  year = {2024},
  month = {Feb},
  publisher = {American Physical Society}
}

@article{heijden+al:2001,
  author = {G. H. M. Van der Heijden and S. Neukirch and V. G. A. Goss and J. M. T. Thompson},
  journal = {International Journal of Mechanical Sciences},
  number = {1},
  pages = {161--196},
  title = {Instability and self-contact phenomena in the writhing of clamped rods},
  volume = {45},
  year = {2003}
}

@book{audoly_elasticity_2005,
  title = {Elasticity and geometry: from hair curls to the non-linear response of shells},
  author = {Basile Audoly and Yves Pomeau},
  year = {2010},
  publisher = {Oxford University Press}
}

@article{coleman+swigon:2000,
  author = {B. D. Coleman and D. Swigon},
  journal = {Journal of Elasticity},
  pages = {173--221},
  title = {Theory of supercoiled elastic rings with self-contact and its application to {DNA} plasmids},
  volume = {60},
  year = {2000}
}

@book{oreilly2017,
  author = {Oliver M. O'Reilly},
  publisher = {Springer},
  title = {Modeling nonlinear problems in the mechanics of strings and rods},
  year = {2017}
}

@article{Singh2022,
  author = {Harmeet Singh},
  journal = {Journal of Elasticity},
  pages = {321--346},
  title = {Planar Equilibria of an Elastic Rod Wrapped Around a Circular Capstan},
  volume = {151},
  year = {2022}
}

@article{Neukirch2025Noetherian,
  author = {S{\'e}bastien Neukirch and Florence Bertails-Descoubes},
  journal = {International Journal of Non-Linear Mechanics},
  pages = {105269},
  title = {A Noetherian approach to invariants for the statics and dynamics of elastic rods},
  year = {2025}
}

@article{baek_finite_2020,
  title = {Finite element modeling of tight elastic knots},
  author = {Changyeob Baek and Paul Johanns and Tomohiko G. Sano and Paul Grandgeorge and Pedro M. Reis},
  journal = {Journal of Applied Mechanics},
  volume = {88},
  number = {2},
  pages = {024501},
  year = {2020},
  month = {Nov}
}

@article{jung_capstan_2008,
  title = {Capstan equation including bending rigidity and non-linear frictional behavior},
  author = {Jae Ho Jung and Ning Pan and Tae Jin Kang},
  journal = {Mechanism and Machine Theory},
  volume = {43},
  number = {6},
  pages = {661--675},
  year = {2008},
  month = {Jun}
}

@book{ashley_ashley_1944,
  title = {The {Ashley} {Book} of {Knots}},
  author = {Clifford Warren Ashley},
  publisher = {Knopf Doubleday Publishing Group},
  year = {1944}
}

@article{jawed_untangling_2015,
	title = {Untangling the mechanics and topology in the frictional response of long overhand elastic knots},
	volume = {115},
	number = {11},
	journal = {Physical Review Letters},
	author = {Jawed, M. K. and Dieleman, P. and Audoly, B. and Reis, P. M.},
	month = sep,
	year = {2015},
	pages = {118302},
}

@article{johanns_shapes_2021,
	title = {The shapes of physical trefoil knots},
	volume = {43},
	issn = {2352-4316},
	journal = {Extreme Mechanics Letters},
	author = {Johanns, Paul and Grandgeorge, Paul and Baek, Changyeob and Sano, Tomohiko G. and Maddocks, John H. and Reis, Pedro M.},
	month = feb,
	year = {2021},
	pages = {101172},
}

@article{sano_exploring_2022,
  title = {Exploring the inner workings of the clove hitch knot},
  author = {Tomohiko G. Sano and Paul Johanns and Paul Grandgeorge and Changyeob Baek and Pedro M. Reis},
  journal = {Extreme Mechanics Letters},
  volume = {55},
  pages = {101788},
  year = {2022},
  month = {Aug}
}

@article{stuart_capstan_1961,
  title = {Capstan equation for strings with rigidity},
  author = {I. M. Stuart},
  journal = {British Journal of Applied Physics},
  volume = {12},
  number = {10},
  pages = {559--562},
  year = {1961}
}

@article{grandgeorge_mechanics_2021,
	title = {Mechanics of two filaments in tight orthogonal contact},
	volume = {118},
	number = {15},
	urldate = {2024-06-05},
	journal = {Proceedings of the National Academy of Sciences},
	author = {Grandgeorge, Paul and Baek, Changyeob and Singh, Harmeet and Johanns, Paul and Sano, Tomohiko G. and Flynn, Alastair and Maddocks, John H. and Reis, Pedro M.},
	month = apr,
	year = {2021},
	pages = {e2021684118},
}

@article{grandgeorge_elastic_2022,
  title = {An elastic rod in frictional contact with a rigid cylinder},
  author = {Paul Grandgeorge and Tomohiko G. Sano and Pedro M. Reis},
  journal = {Journal of the Mechanics and Physics of Solids},
  volume = {164},
  pages = {104885},
  year = {2022},
  month = {Jul}
}

@book{eytelwein1842handbuch,
  title = {Handbuch der Mechanik fester K{\"o}rper und der Hydraulik},
  author = {Johann Albert Eytelwein},
  publisher = {Fleischer},
  year = {1842}
}

@article{euler_remarques_1769,
  title = {Remarques sur l'effet du frottement dans l'équilibre},
  author = {Leonhard Euler},
  journal = {Mémoires de l'académie des sciences de Berlin},
  pages = {265--278},
  year = {1769},
  month = {Jan}
}

@article{johanns_strength_2023,
  title = {The strength of surgical knots involves a critical interplay between friction and elastoplasticity},
  author = {Paul Johanns and Changyeob Baek and Paul Grandgeorge and Selim Guerid and Stephen A. Chester and Pedro M. Reis},
  journal = {Science Advances},
  volume = {9},
  number = {23},
  pages = {eadg8861},
  year = {2023}
}

@article{johanns_capsizing_2024,
  title = {Capsizing due to friction-induced twist in the failure of stopper knots},
  author = {Paul Johanns and Pedro M. Reis},
  journal = {Extreme Mechanics Letters},
  volume = {68},
  pages = {102134},
  year = {2024}
}

@article{audoly_elastic_2007,
  title = {Elastic {Knots}},
  author = {Basile Audoly and Nicolas Clauvelin and S{\'e}bastien Neukirch},
  journal = {Physical Review Letters},
  volume = {99},
  number = {16},
  pages = {164301},
  year = {2007},
  month = {Oct}
}

@article{aymon2025self,
  title = {Self-locking and stability of the bowline knot},
  author = {Bastien F. G. Aymon and Fani Derveni and Michael Gomez and J{\'e}r{\^o}me Crassous and Pedro M. Reis},
  journal = {Extreme Mechanics Letters},
  volume = {81},
  pages = {102413},
  year = {2025}
}

@article{day1986art,
  title = {The art of knotting and splicing},
  author = {Cyrus Lawrence Day},
  journal = {Naval Institute Press},
  year = {1986}
}

@article{wright1928knots,
  title = {Knots for climbers},
  author = {CEI Wright and JE Magowan},
  journal = {Alpine Journal},
  volume = {40},
  number = {120},
  pages = {340},
  year = {1928}
}

@article{zimmer1991influence,
  title = {Influence of knot configuration and tying technique on the mechanical performance of sutures},
  author = {Christopher A. Zimmer and John G. Thacker and David M. Powell and Kenneth T. Bellian and Daniel G. Becker and George T. Rodeheaver and Richard F. Edlich},
  journal = {The Journal of Emergency Medicine},
  volume = {9},
  number = {3},
  pages = {107--113},
  year = {1991}
}

@article{kim2001significance,
  title = {Significance of the internal locking mechanism for loop security enhancement in the arthroscopic knot},
  author = {Seung-Ho Kim and Kwon-Ick Ha and Sang-Hyun Kim and Jung-Sung Kim},
  journal = {Arthroscopy: The Journal of Arthroscopic \& Related Surgery},
  volume = {17},
  number = {8},
  pages = {850--855},
  year = {2001}
}

@article{seguin2022twist,
  title = {Twist-controlled force amplification and spinning tension transition in yarn},
  author = {Antoine Seguin and J{\'e}r{\^o}me Crassous},
  journal = {Physical Review Letters},
  volume = {128},
  number = {7},
  pages = {078002},
  year = {2022}
}

@article{warren2018clothes,
  title = {Why clothes don’t fall apart: Tension transmission in staple yarns},
  author = {Patrick B. Warren and Robin C. Ball and Raymond E. Goldstein},
  journal = {Physical Review Letters},
  volume = {120},
  number = {15},
  pages = {158001},
  year = {2018}
}

@article{poincloux2018crackling,
  title = {Crackling dynamics in the mechanical response of knitted fabrics},
  author = {Samuel Poincloux and Mokhtar Adda-Bedia and Fr{\'e}d{\'e}ric Lechenault},
  journal = {Physical Review Letters},
  volume = {121},
  number = {5},
  pages = {058002},
  year = {2018}
}

@incollection{bueno2019structure,
  title = {Structure and mechanics of knitted fabrics},
  author = {Marie-Ange Bueno and Brigitte Camillieri},
  booktitle = {Structure and Mechanics of Textile Fibre Assemblies},
  publisher = {Woodhead Publishing},
  pages = {61--107},
  year = {2019}
}

@article{goriely2006mechanics,
  title = {Mechanics of climbing and attachment in twining plants},
  author = {Alain Goriely and S{\'e}bastien Neukirch},
  journal = {Physical Review Letters},
  volume = {97},
  number = {18},
  pages = {184302},
  year = {2006}
}

@article{silk2005importance,
  title = {The importance of frictional interactions in maintaining the stability of the twining habit},
  author = {Wendy K. Silk and N. Michele Holbrook},
  journal = {American Journal of Botany},
  volume = {92},
  number = {11},
  pages = {1820--1826},
  year = {2005}
}

@article{childs2019belt,
  title = {Belt and chain drives},
  author = {P. R. N. Childs},
  journal = {Mechanical Design Engineering Handbook},
  pages = {533--597},
  year = {2019},
  publisher = {Elsevier}
}

@article{baser2010theoretical,
  title = {Theoretical and experimental determination of capstan drive slip error},
  author = {Ozgur Baser and E. Ilhan Konukseven},
  journal = {Mechanism and Machine Theory},
  volume = {45},
  number = {6},
  pages = {815--827},
  year = {2010}
}

@article{weiner2020mechanics,
  title = {Mechanics of randomly packed filaments—the “bird nest” as meta-material},
  author = {Nicholas Weiner and Yashraj Bhosale and Mattia Gazzola and Hunter King},
  journal = {Journal of Applied Physics},
  volume = {127},
  number = {5},
  pages = {050902},
  year = {2020}
}

@article{hertz1881contact,
  title={The contact of elastic solids},
  author={Hertz, Heinrich},
  journal={J Reine Angew, Math},
  volume={92},
  pages={156--171},
  year={1881}
}

\end{document}